# Nanoscale rheology: Dynamic Mechanical Analysis over a broad and continuous frequency range using Photothermal Actuation Atomic Force Microscopy

Alba R. Piacenti[a], Casey Adam[a,b], Nicholas Hawkins[b], Ryan Wagner[c], Jacob Seifert[a], Yukinori Taniguchi[d], Roger Proksch[e], Sonia Contera[a]*

[a]Clarendon Laboratory, Department of Physics, University of Oxford, OX1 3PU Oxford, United Kingdom

[b] Department of Engineering Science, University of Oxford, OX1 3PJ Oxford, United Kingdom

[c]School of Mechanical Engineering, Purdue University, West Lafayette, Indiana, 47907, United States

[d]Asylum Research, Oxford Instruments KK, Tokyo 103-0006, Japan

[f]Asylum Research-An Oxford Instruments Company, Santa Barbara, California 93117, United States

* Corresponding Author; E-mail: sonia.antoranzcontera@physics.ox.ac.uk

## Abstract

Polymeric materials are widely used in industries ranging from automotive to biomedical. Their mechanical properties play a crucial role in their application and function, and arise from the nanoscale structures and interactions of their constitutive polymer molecules. Polymeric materials behave viscoelastically, i.e. their mechanical responses depend on the time scale of the measurements; quantifying these time-dependent rheological properties at the nanoscale is relevant to develop, for example, accurate models and simulations of those materials, which are needed for advanced industrial applications. In this paper, an atomic force microscopy (AFM) method based on the photothermal actuation of an AFM cantilever is developed to quantify the nanoscale loss tangent, storage modulus, and loss modulus of polymeric materials. The method is then validated on a styrene-butadiene rubber (SBR), demonstrating the method's ability to quantify nanoscale viscoelasticity over a continuous frequency range up to five orders of magnitude (0.2 Hz to 20,200 Hz). Furthermore, this method is combined with AFM viscoelastic mapping obtained with amplitude-modulation frequency-modulation (AM-FM) AFM, enabling the extension of viscoelastic quantification over an even broader frequency range, and demonstrating that the novel technique synergizes with preexisting AFM techniques for quantitative measurement of viscoelastic properties. The method presented here introduces a way to characterize viscoelasticity of polymeric materials, and soft matter in general at the nanoscale, for any application.

## INTRODUCTION

Polymeric materials are widely used in many different types of applications, and exhibit time- and frequency- dependent mechanical behavior known as viscoelasticity[1]. The viscoelastic properties of polymeric materials are crucial to their function and application, and arise from the structure and interactions of polymers within the material[1,2]. Quantifying material viscoelasticity is therefore essential in determining the material's application, and in providing insight into the material's structure[1,2]. Typically, viscoelasticity is quantified at the macroscale, using techniques such as dynamic mechanical analysis (DMA) or rheometry[1,3]. However, it is also useful (e.g. to construct or validate predictive models of polymeric behavior) though more technically demanding, to quantify viscoelasticity at the nanoscale, since this is the length scale at which polymers interact within the material. In both macro and nano-DMA or rheology, an axial or torsional stimulus is applied to a whole sample. For macroscale measurements, a large, typically mm, stimulus is applied to the sample[1,3]. For nano-DMA, the stimulus is applied to a localized position on the sample, typically nm or µm in size, by a nanoscale or microscale size probe; some methods also stimulate the whole sample, and measure its response locally, using the probe [4,5].

Among techniques to quantify the nanoscale viscoelasticity of polymeric materials, atomic force microscopy (AFM) is one of the most versatile. In AFM, it is possible to apply a wide range of forces, from pN to µN, to a sample and probe samples at different length scales, from nm to hundreds of µm, depending on the stiffness, tip size, and shape of the AFM cantilever[6,7]. Moreover, localized AFM measurements can be combined to create quantitative maps of a sample's mechanical properties with high spatial resolution[7]. Furthermore, AFM can be used in liquid and at different temperatures[6], allowing inert materials and biological (even living) samples to be measured in conditions similar to those of their application. Lastly, in general the AFM requires no external fields that might interfere with the natural behavior of the studied material.

Several AFM techniques, including contact resonance (CR) and multifrequency AFM, have been used to map the viscoelastic properties of samples at frequencies corresponding to the AFM cantilever's harmonics or eigenmodes[8–16]. However, measuring properties over a wide frequency range is preferred because sample viscoelasticity is frequency dependent. The wider the measured frequency range, the more is known about a material's viscoelastic behavior and its relation to the internal molecular structure. Furthermore, following polymer physics theory, the time scales that are relevant for polymers at the nanometre scale are much smaller than those at the macroscopic scale, hence nano-mechanical techniques need to reach the kHz range to be physically meaningful. In recent years, off-resonance AFM nano/micro-rheology has been developed to study the viscoelastic properties of many different materials, including rubbers[17–22], cells[23–27], single cell nuclei[28], cartilage[29–33], and polymer gels[23,34]. However, there are limitations shared by existing AFM nano/micro-rheology techniques. The first limitation is that the frequencies over which properties can be measured are limited by reliance on piezoelectric (PE) actuators to excite the cantilever. PE actuators can introduce spurious peaks in the cantilever's oscillatory spectrum, especially in liquid,[35–37] and thereby causing noise in rheological measurements, and rendering experiments unreliable or difficult to analyze (especially on biological samples). So far, different solutions have been used to overcome the limited frequency range of PE-actuated systems, including adaptation of high-frequency piezo actuators[18–20,26], compensation for PE resonances[27], application of the time-temperature superposition (TTS) principle[17,18,21,22], or using direct cantilever excitation via magnetic actuation[34]. Nevertheless, the spurious spectrum emerging from unwanted resonances still limits PE methods, and TTS or magnetic actuation might alter sample behavior by compromising the material via temperature change (e.g. leading to DNA/biomolecular denaturation), application of magnetic fields (e.g. magneto-active materials) and bio-toxicity of magnetic coatings of cantilevers.

Photothermal (PT) actuation is another way of directly exciting AFM cantilevers[38,39], and is already used in commercial instruments to quantify the mechanical properties of materials both on-resonance[15,40], and off-resonance[41,42], and is particularly useful for biological samples in liquid environments[16]. In this paper, we develop a nano/microscale rheology AFM technique using PT cantilever actuation and show that this method can accurately measure the viscoelastic properties of a sample over a continuous and wide frequency range five orders in magnitude.

## RESULTS AND DISCUSSION

**PT-AFM nano-DMA.** The principle of our PT-AFM nano-DMA technique is shown in Figure 1. PT cantilever excitation is achieved by modulating the power of an excitation laser (EL). The PT-EL is focused on the back of the cantilever and drives cantilever motion by PT excitation[38,39]. To obtain a continuous wide spectrum of frequencies, the PT-EL is modulated with exponentially chirped oscillations at frequencies well below the cantilever's resonance[41]. Details on why exponential chirps were used as well as analysis of the effects of PT-EL power and positioning, are provided in the Supporting Information (SI3 and SI4). Comparing chirped cantilever oscillations above (Figure 1(a)) and during indentation (Figure 1(b)) of the sample allows quantification of sample viscoelasticity. The measurement above the sample while the cantilever is out of contact acts as a reference measurement. The measurement performed while indenting, and being in contact with the sample, is the sample measurement. A detailed analysis of factors that can influence cantilever motion (e.g. PT-EL position, PT-EL displacements during cantilever motion, and the extent by which tip/sample separation affects the reference signal) and sample response (e.g. non-linear viscoelastic behavior due to high strains caused by large oscillations or indentations), and hence influence PT-AFM nano-DMA measurements, is provided in the Supporting Information (SI4), and shows that PT-AFM nano-DMA is robust to most of these factors, and reliably quantifies sample viscoelasticity.

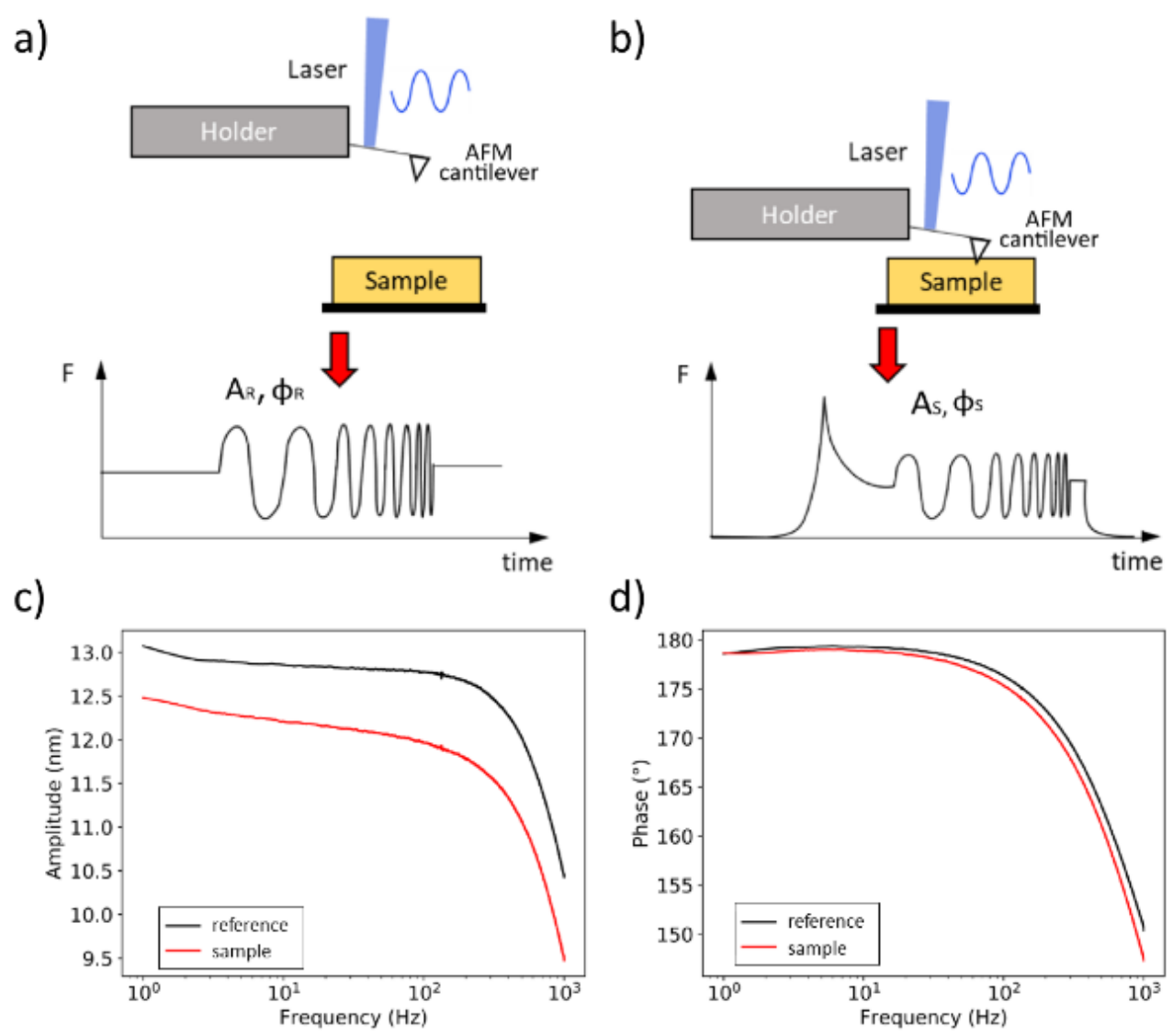

Figure 1. Principle of PT-AFM nano-DMA (photothermal atomic force microscopy nano-dynamic mechanical analysis). A reference measurement, where the cantilever is excited by a chirped oscillation while not in contact with the sample, is shown in (a). A sample measurement, where the cantilever is excited by the same chirped oscillation while in contact with the sample, is shown in (b). Schematics of the force (F) experienced by the cantilever are shown at the bottom of (a) and (b). Cantilever amplitude (A) and phase (φ) vary between the reference (R) and sample (S) measurements. Representative amplitude and phase changes between sample (a styrene-butadiene rubber (SBR), red) and reference (black) measurements are shown in (c) and (d), respectively. Sample viscoelasticity is calculated by comparing $A_R$ and $\varphi_R$ with $A_S$ and $\varphi_S$.

Figure 1(c, d) shows representative amplitude and phase signals of chirped AFM cantilever (an AC160-TSA) oscillations during a reference measurement and during indentation of a styrene-butadiene rubber (SBR) sample. For both measurements, amplitude and phase decrease with increasing frequency, which is typical of PT measurements[43]. The differences between the amplitude and phase between the sample and reference measurements are used to calculate sample viscoelasticity. As derived by Nalam et *al.* [34], the real (k') and imaginary (k'') components of the dynamic stiffness of a sample (k*=k'+ik'') probed with sinusoidally directly actuated cantilevers can be calculated from amplitude (A) and phase ($\varphi$) as follows:

$$k' = k_c(\bar{A}\cos\Phi - 1) \quad (1)$$

$$k'' = k_c\bar{A}\sin\Phi \quad (2)$$

Where $\bar{A} = A_R/A_S$ and $\Phi = \varphi_R - \varphi_S$, with A and $\varphi$ being the amplitude and phase of the oscillations measured out of contact (reference measurement, subscript R) and in contact (sample measurement, subscript S) with the sample.

To directly apply Eq. 1 and Eq 2. to PT-AFM nano-DMA, the optical lever calibration of both $A_R$ and $A_S$ must be identical. This calibration depends on the shape in which the cantilever vibrates, which generally changes with photothermal laser spot position, drive frequency, and cantilever boundary conditions [41, 57] (these factors are analyzed further in supplementary material section SI4). Without correcting for these shape changes, measurements at different ratios of k*: $k_c$ are not directly comparable. However, if $k_c$ >> k* the change in cantilever vibration shape due to changes in sample stiffness is small [41]. Selecting k*: $k_c$ ratios that are sensitive to amplitude changes but insensitive to vibration shape changes, and comparing $A_R$ and $A_S$ at matching frequencies are therefore essential for PT-AFM nano-DMA to provide accurate results. Examples that illustrate what happens when these conditions are violated are demonstrated in the Supporting Information SI4.

The loss tangent, tan $\delta$, of a sample can be calculated as the ratio of the imaginary and the real components of k* as follows[19]:

$$\tan\delta = \frac{k''}{k'} \quad (3)$$

Importantly, contrary to k' and k'', tan $\delta$ does not depend on the geometry of the system[3], in this case the tip/sample contact. Calculation of the measurement uncertainty in k' and k'' can be found in the Supporting Information (SI5).

Figure 2 (a) shows SBR k' and k'' calculated using Eqs. 1 and 2. Both values increase with increasing frequency, with k'' becoming larger than k' above 2 kHz; tanδ was then calculated using Eq. 3 and compared with control SBR measurements obtained via macroscale DMA (figure 2 (b)). PT-AFM nano-DMA measurements are comparable with macroscopic DMA data at 42°C rather than 30°C (ambient room temperature), suggesting a local increase of sample temperature during PT-AFM nano-DMA measurements. This local heating is likely due to the PT-EL as further described in the Supporting Information (SI4). At frequencies less than 1 Hz, tanδ measured with PT-AFM nano-DMA deviated slightly from macroscale DMA measurements. However, small deviations between local and macroscopic rheological measurements have been previously reported[19,21,34], in some cases attributed to non-linear viscoelastic effects[21]. However, as demonstrated in the Supporting Information (SI4), nonlinear effects are avoided in our SBR measurements. More likely, the deviation between local and macroscopic DMA measurements of the SBR are due to surface effects that come into play at smaller length scales[34]. Regardless of these differences, macroscopic measurements could potentially miss features relevant at smaller length scales by averaging over a larger scale, furthermore the bulk oscillation of the whole sample is also likely to affect the local mechanical properties in a different way than nanoscale oscillations confined to the place of the measurement. Small deviations at high frequencies (approximately 10-20 kHz) could similarly be due to surface effects, although |k*| approaching the value of $k_c$ might also be affecting the measurements (see Supporting Information SI4).

To evaluate how PT measurements compare with PE measurements, AFM nano-DMA measurements of the SBR were obtained using PE actuation via sample modulation, and compared to PT-AFM nano-DMA measurements. Additionally, PE nano-DMA measurements were used to provide insight into the local temperature increase caused by the PT-EL. Cantilever excitation was chosen to keep similar reference amplitudes for PE and PT measurements (see Supporting Information SI1). PE and PT measurements were performed on the same spot of the SBR sample. PE measurements were performed first (i), followed by PT measurements (ii). Then, PE measurements were performed keeping the PT-EL focused on the cantilever with only DC power (iii), that is with the PT-EL but without any applied PT-EL oscillation. Finally, the PT-EL was deactivated, and PE measurements were performed a second time on a different spot on the sample (iv). The relationships obtained by Igarashi *et al.*[19] were used to calculate k' and k'' from PE measurements. As shown in the Supporting Information (SI1), data obtained with PE actuation could not be measured above 1 kHz due to the presence of spurious resonances, arising from the PE actuator-AFM coupled system[35–37]; tanδ values measured by (i-iv) are shown in Figure 3. Unlike PT data (ii), PE data (i) agree well with macroscopic DMA at 30°C. PE data obtained when the PT-EL was focused on the cantilever, but not used to excite the cantilever (iii), matches DMA data at the intermediate temperature of 34°C. After switching off the PT-EL, another PE measurement was performed (iv), and the data once again overlap well with DMA at 30°C. These observations from (i)-(iv) indicate that the PT-EL causes local sample heating. Nevertheless, it would be worth investigating this effect further either using thermocouples in proximity of the sample/cantilever or using AFM cantilevers with integrated thermometers[44,45].

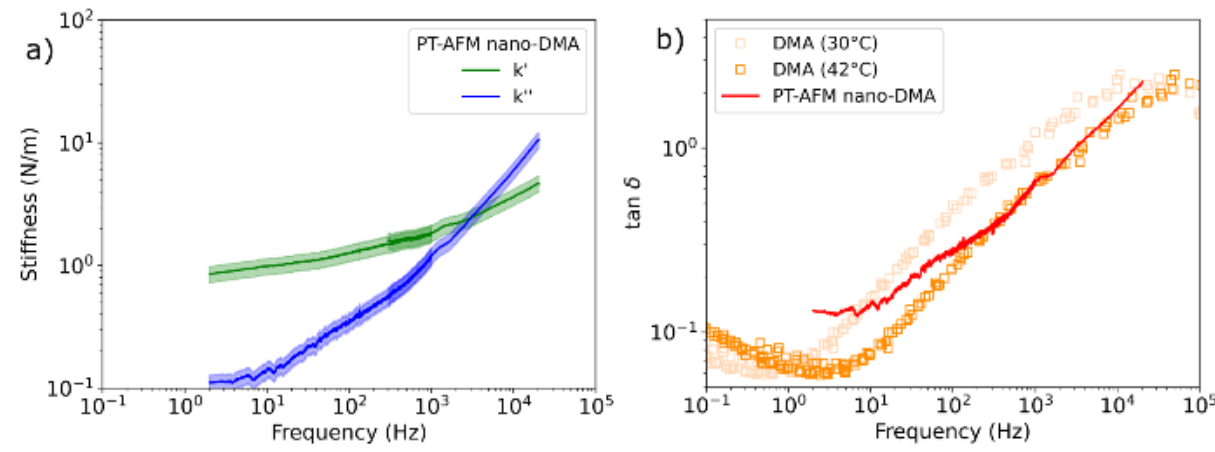

Figure 2. SBR k', k'' and tanδ measured using PT-AFM nano-DMA. The storage stiffness (k', green line) and loss stiffness (k'', blue line) are shown in (a). Comparisons between the loss tangent (tanδ) measured with PT-AFM nano-DMA (red line) and macroscopic DMA (unfilled squares) master curves with reference temperatures of 42° (dark orange) and 30° (peach) are shown in (b). Shading represents the experimental error, calculated as detailed in the Supporting Information (SI5), of PT-AFM nano-DMA measurements.

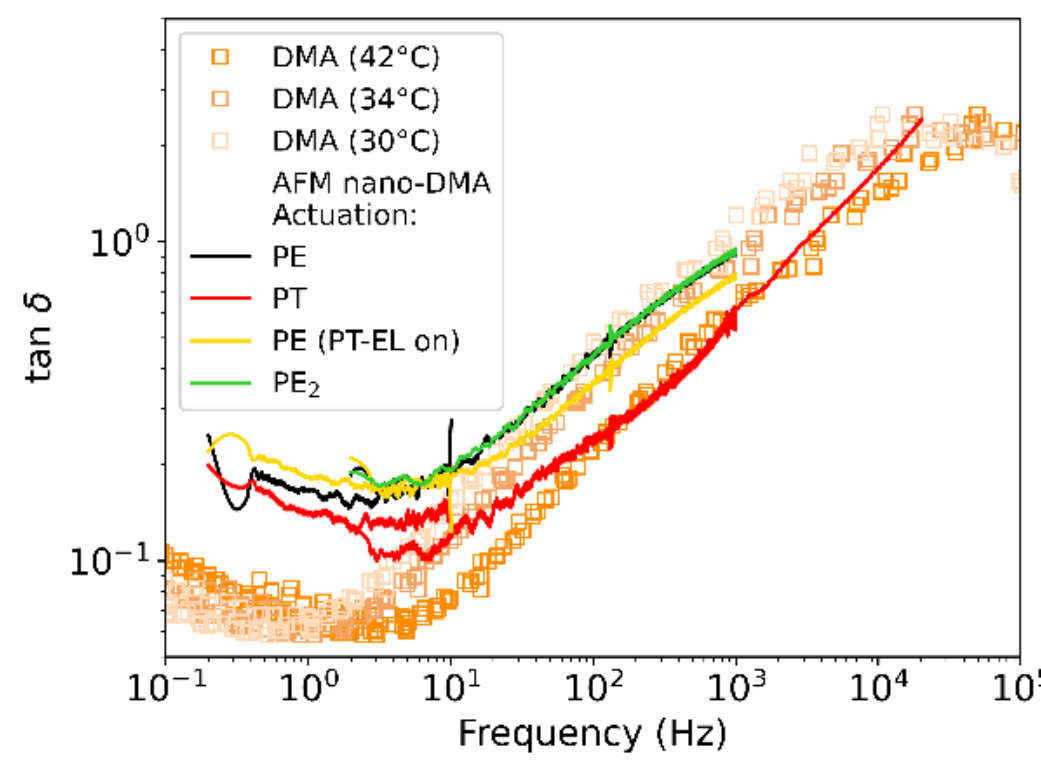

Figure 3. Evaluation of PT-AFM nano-DMA against PE-AFM nano-DMA and macroscale DMA measurements of the SBR. AFM nano-DMA measurements were obtained with PE (black line) and PT (red line) excitation. PE measurements were performed first. Then PT measurements were performed. Next, PE measurements were performed while the PT-EL was still focused on, but not exciting, the cantilever (yellow line). Finally, PE measurements were repeated after deactivating the PT-EL (green line). Macroscopic DMA (unfilled squares) master curves with reference temperatures of 42°C (dark orange), 34°C (light orange) and 30°C (peach) are also shown as controls for nano-DMA measurements.

**Dynamic modulus.** Typically, viscoelastic materials are not described in terms of k' and k'', but instead by their dynamic modulus (E*=E'+iE'')[1]. The real and imaginary components of E* are called the storage (E') and the loss (E'') moduli, and respectively represent the elastically stored energy density and the energy density dissipated during sample deformation[1,3]. E* can be calculated from k* by applying a contact model to describe the indenter-sample system. The presence of adhesive forces can complicate the contact model equations[19]. For dynamic experiments on adhesive viscoelastic materials, as long as the oscillation frequency is high enough, viscoelastic effects cause indenters of different shapes to behave like a flat cylindrical punch (i.e. constant contact radius) during the oscillations [46,47]. For these cases, the relation between E* and k* is written as[46,47]:

$$E^* = \frac{1-\nu_S^2}{2a}k^* \tag{4}$$

Where $a$ is the contact radius between the indenter (I) and the sample (S) during the oscillations, and $\nu_S$ is the Poisson's ratio of the sample. In this paper, $\nu_S = 0.5$ was assumed, which is typical of SBRs[19–22]. From Eq. 4, it can also be seen that, for a relationship where E* and k* are directly proportional, Eq. 3 results in tanδ = k''/k' = E''/E', which is the usual definition of tanδ[1-3].

To calculate $a$ it is necessary to determine the correct contact model to describe the system. For soft and adhesive rubbery materials like the SBR, it is appropriate to use contact models such as the Johnson-Kendall-Robertson (JKR) model[19], originally introduced for deformation of spherical bodies[48]. In general, the following relationships can be used to describe the relationship between the force F exerted by a rigid indenter and the deformation d of an elastic half space[49]:

$$d(a) = d_{NA} - \sqrt{\frac{2\pi w a}{\tilde{E}}} \tag{5}$$

$$F(a) = F_{NA} - \sqrt{8\pi a^3 w \tilde{E}} \tag{6}$$

Where $d_{NA}$ and $F_{NA}$ are the deformation and the force calculated in the non-adhesive case, $w$ is the energy of adhesion per unit contact area, and $\tilde{E}$ is the reduced Young's modulus defined as[6]: $\frac{1}{\tilde{E}} = \frac{1-\nu_S^2}{E_S} + \frac{1-\nu_I^2}{E_I} \approx \frac{1-\nu_S^2}{E_S}$, if $E_I \gg E_S$ (as usual in AFM experiments).

For indenters with hyperboloid shapes, Sun *et al.*[50] proposed a model that can be simplified to the following (details in Supporting Information SI2):

$$d = \frac{aA\pi}{2R} - \sqrt{\frac{2\pi w a}{\tilde{E}}} \tag{7}$$

$$F = \frac{2\tilde{E}A}{2R}\left[aA + \frac{a^2 - A^2}{2}\pi\right] - \sqrt{8\pi a^3 w \tilde{E}} \tag{8}$$

Where A = R cotα, with α being the indenter semi-vertical angle and R the tip radius. For AC160 and AC240 cantilevers, R was assumed to be the nominal cantilever radius (R = 7 nm) and α was taken as half of the tip's nominal back angle (α = 17.5°). Note that, while the tips used in our experiments have a pyramidal geometry, as described by the manufacturer, it is, however, impossible to measure the actual geometry at the nanoscale contact. The hyperboloidal model has been shown to be a good approximation (Eq. 7,8) of the contact geometry of our system, as demonstrated by the fitting to the experimental data presented in Fig. 4, and in good agreement with macorscopic DMA measurements.

To calculate $w$ and $\tilde{E}$ necessary to obtain $a$, and therefore E' and E'', the "*two-points method*"[50], which relies on particular points in force indentation withdraw curves, can be used[19]. In Figure 4, the relevant points are marked in a typical force indentation curve obtained for a PT-AFM nano-DMA experiment on the SBR: the point of zero load "0", the point around which dynamic oscillations occurr "1"[19], and the point of zero deformation "2". Figure 4 also compares the measured withdraw curve to the curves for the spherical JKR (Eq. 5,6) and hyperboloid (Eq. 7,8) contact models evaluated using the quantities calculated with the "*two-points method*"[50] (see Supporting Information SI2). The hyperboloid model agrees well with the experimental curve for larger indentations, where the model assumpion that $a$ >>A (see Supporting Information SI2) is more likely to be satisfied.

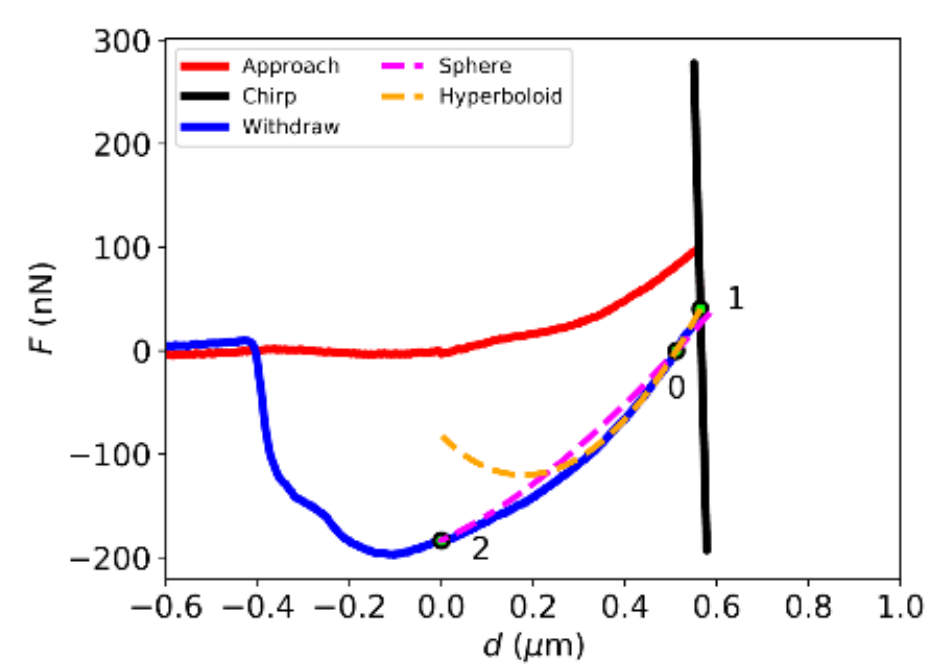

Figure 4. Typical force (F) versus indentation (d) curve of the SBR from a PT-AFM nano-DMA experiment using an AC160 cantilever and a chirp frequency range of 0.1 Hz to 10 Hz. The approach (red line), chirped oscillations (black line), and withdraw (blue line) curves are shown. Point "0" corresponds to the point of zero load, "1" to the average of the oscillatory force (i.e. the point around which dynamic oscillations occur), and "2" to the point of zero indentation. Dashed lines represent contact model fits, evaluated by calculating $w$ and $\tilde{E}$, for a spherical JKR contact geometry (pink line) and a hyperboloid (orange line) contact geometry.

For PT-AFM nano-DMA, calculating E' and E'' (*Eq.4*) requires knowledge of $a$ in point "1" ($a_1$). The value of $a_1$ was calculated for a spherical, conical, and hyperboloid indenter once $w$ and $\tilde{E}$ were obtained for each geometry, as detailed in the Supporting Information (SI2). The procedure to calculate the uncertainty in E' and E'' is described in the Supporting Information (SI6). The resulting SBR E' and E'' are shown as lines in Figure 5(a,b), and compared to the macroscopic DMA control (unfilled squares). E' and E'' were overestimated by PT-AFM nano-DMA when $a_1$ was calculated via the spherical and conical indenter models, most likely because the contact geometry is not well described by these two models. E' and E'' calculated using $a_1$ calculated with the hyperboloid contact model are in good agreement with macroscopic DMA data. Slight deviations between moduli measured with PT-AFM nano-DMA and macroscopic DMA at low frequencies can be explained as proposed above for tanδ measurements.

It is important to note that other contact models could also be employed to describe the SBR tip/sample contact. For example, the indentation curve shown in Figure 4 potentially exhibits characteristics of plastic deformation[6]. However, since multiple F vs. d curves performed in succession exhibited the same shape (shown in Figure 4) which would not be the case for plastic deformation, since the hyperboloid with adhesion fits the F vs. d curves well (Figure 4), and since the resulting SBR E' and E'' match control DMA measurements (Figure 5), it is reasonable to conclude that plastic deformation did not significantly impact PT-AFM nano-DMA measurements of this sample.

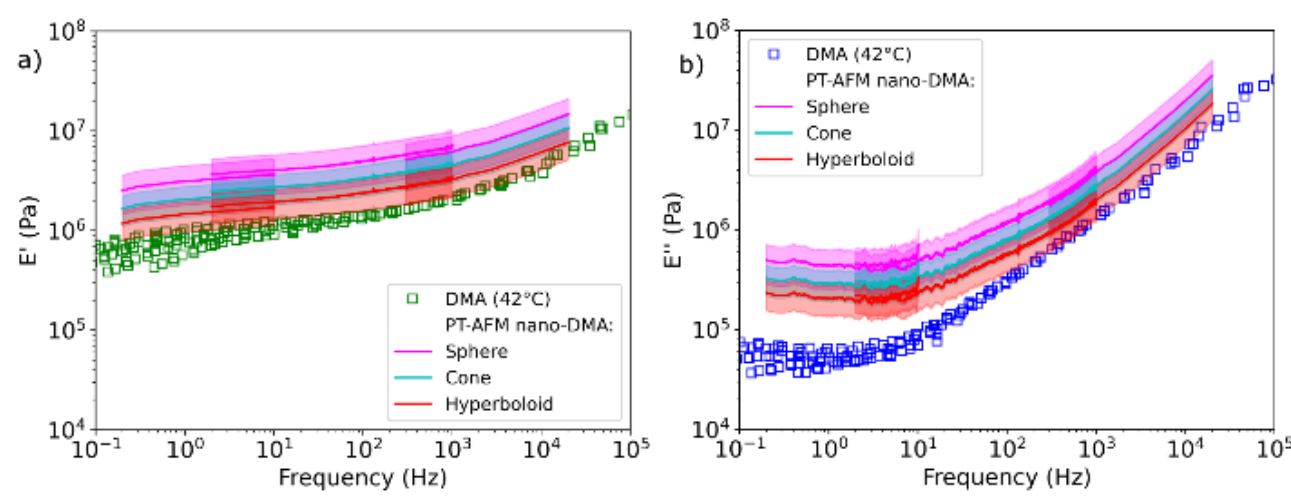

Figure 5. SBR storage (E') and loss (E'') moduli measured by PT-AFM nano-DMA (continuous lines) using different contact models. (a) E', and (b) E'' were calculated from k' and k'' via Eq.4 for adhesive contacts with spherical (pink line), conical (cyan line), and hyperboloid (red line) indenters (see Supporting Information SI2). Measurement errors (see Supporting Information SI6) are shown as shades. Unfilled squares represent the macroscopic DMA control measurement at 42°C.

**AM-FM AFM Imaging**. In order to evaluate how PT-AFM nano-DMA measurements compare to other AFM techniques that are used to measure viscoelastic properties of soft materials, the SBR sample was also measured using Amplitude Modulation-Frequency Modulation (AM-FM) AFM. AM-FM AFM is an on-resonance technique that measures sample viscoelasticity by simultaneously driving the cantilever at two of its eigenmodes, typically the first and second[14]. AM-FM AFM allows quantitative mapping of sample topography, tanδ, and E'[13–15]. The first, lower-frequency, mode is subject to amplitude modulation, and measures sample topography and tan $\delta$[51,52]. The second, higher-frequency, mode is subject to frequency modulation, and, combined with parameters from the first mode, measures sample E' by applying the Hertz contact model[14,15]. While both modes contribute to the calculation of E', the second mode contributes most[14]. Hence AM-FM AFM measures sample tanδ at the lower frequency, and E' at the higher frequency[14,15,56]. Figure 6(a-c) shows representative AM-FM AFM maps of the SBR. The images in Figure 6(a-c) were obtained using an Olympus AC240 cantilever (nominal $k_{c,1}$ ~2 N/m, $k_{c,2}$ ~ 50 N/m, first resonance frequency ~ 70 KHz, second resonance frequency ~400 kHz), but similar results were obtained using an Olympus AC160 cantilever (nominal $k_{c,1}$ ~26 N/m, $k_{c,2}$ ~ 364 N/m, first resonance frequency ~ 300 KHz, second resonance frequency ~1.5 MHz). Spatial variation was present in SBR topography, tanδ, and E' throughout the SBR surface, as indicated by the color schemes in figure 6(a-c). Mean AM-FM AFM values, shown in Figure 6(d,e), were calculated by averaging all pixels from multiple AM-FM AFM images acquired using AC240 (25 images total) and AC160 (13 images total) cantilevers. These mean AM-FM AFM values are then compared with PT-AFM nano-DMA performed with an AC240 cantilever, and macroscale DMA control measurements. In Figure 6(e), PT-AFM nano-DMA E' was calculated via Eq. 4 for a non-adhesive contact with a spherical indenter (Hertz contact model[53,54], pink line) in order to apply the same contact model as AM-FM AFM measurements, as well as an adhesive contact with a hyperboloid indenter (cyan line, see Supporting Information SI2).

AM-FM AFM tanδ measured with both AC240 and AC160 cantilevers agreed well with control macroscale DMA measurements (Figure 6(b,d)). However, AM-FM AFM E' measured with both cantilevers deviated from control DMA measurements (Figure 6(c,e)). The fact that tanδ is independent of the contact geometry, and did not deviate from the DMA control, while E' relies on a contact model and deviated from the control suggests that the discrepancy between AM-FM AFM and macroscale DMA E' is due to application of an inaccurate contact model to calculate AM-FM AFM E'. Supporting this conclusion, PT-AFM nano-DMA also overestimates SBR E' compared to DMA control measurements when the Hertz model is used. PT-AFM nano-DMA E' calculated using the adhesive hyperboloid model agrees well with macroscale DMA data, as was also the case with the AC160 in Figure 5(a). Therefore, an adhesive hyperboloid contact best describes the AC160/AC240 tip-SBR interaction, and the Hertz contact model used to calculate AM-FM AFM E'[14,15] does not describe the tip/SBR interaction well. The discrepancy between AM-FM AFM and macroscale DMA E' is due to application of an oversimplifying contact model for AM-FM AFM calculations.

Figure 6 also shows that AM-FM AFM tanδ and E' were more sensitive to surface features than PT-AFM nano-DMA, as indicated by the blue squares in Figure 6(a-c). This sensitivity is due to the fact that AM-FM AFM indentation depths are much smaller, a few nm, than AFM nano-DMA indentations (hundreds of nm, as shown in Figure 4), and therefore exhibit greater sensitivity to local properties and surface forces[15,52]. Accounting for AM-FM AFM's application of the simple Hertz contact model when measuring E', and in spite of AM-FM AFM's heightened sensitivity to surface effects, average AM-FM AFM tanδ and E' values agreed with control DMA values and aligned with PT-AFM nano-DMA curves. Therefore, employing both AM-FM AFM and PT-AFM nano-DMA together enhances the amount of information obtained about the sample by measuring sample viscoelasticity over a larger frequency range than either technique alone while also acquiring information about sample topography. If sample topography is varied, performing AM-FM AFM before PT-AFM nano-DMA could also inform where to target PT-AFM nano-DMA measurements on the sample surface. Together, these results demonstrate that PT-AFM nano-DMA measurements synergize well with other AFM techniques that measure sample viscoelasticity.

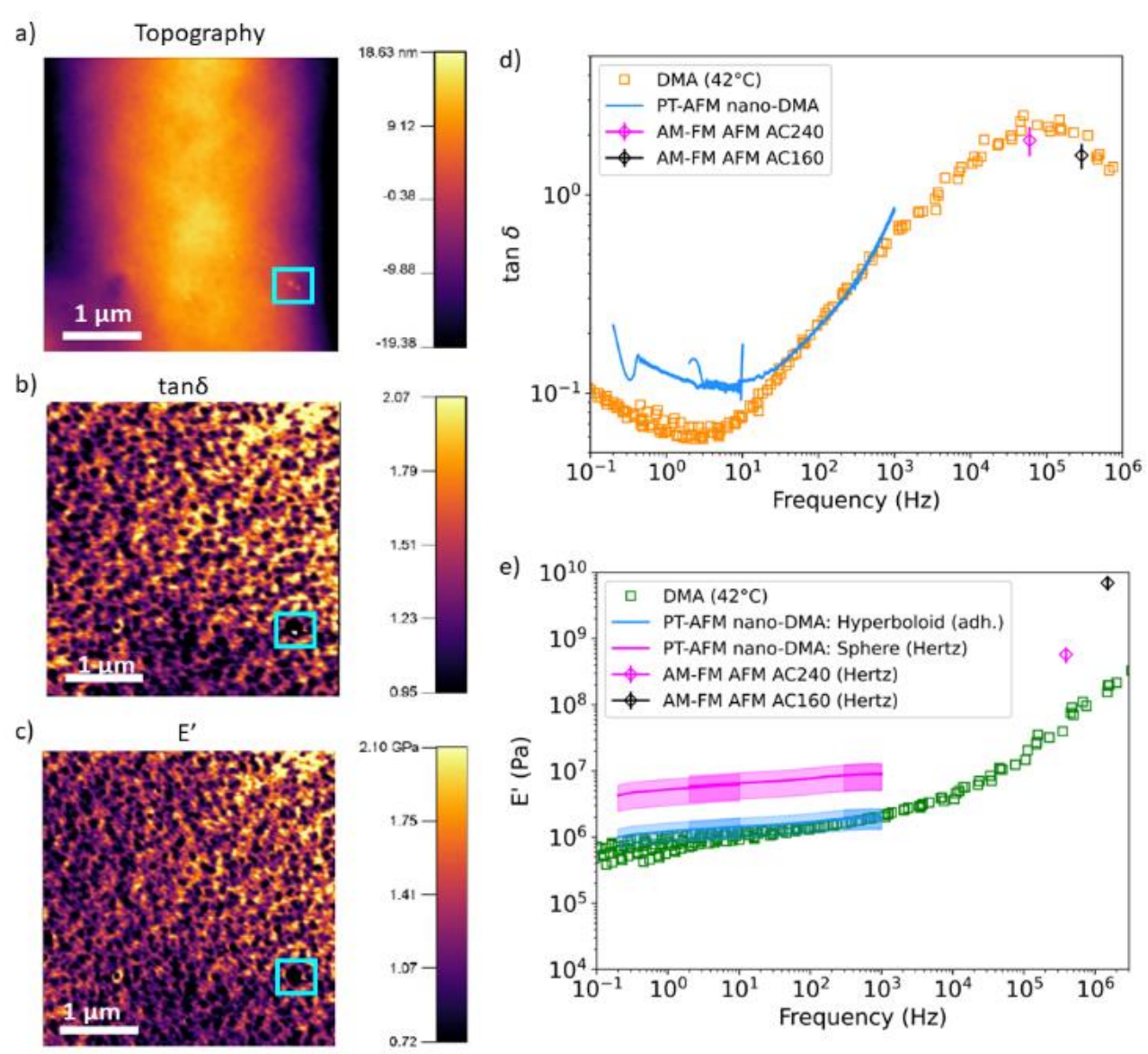

Figure 6: Synergy of PT-AFM nano-DMA and AM-FM AFM measurements. Representative AM-FM AFM maps of SBR topography, loss tangent (tanδ), and storage modulus (E') are shown in (a-c), respectively. Blue squares in (a-c) indicate how SBR surface features affect tanδ and E' values measured by AM-FM AFM. These images were collected with an AC240TSA cantilever, but are representative of all AM-FM AFM images of the SBR surface regardless of cantilever. Comparisons of tanδ and E' values measured by different techniques are respectively shown in (d) and (e). Control macroscopic DMA data (42°C) are displayed as unfilled squares. AM-FM AFM measurements collected with AC240 (pink) and AC160 (black) cantilevers are shown as points with error bars, representing the mean ± standard deviation. PT-AFM nano-DMA data collected with an AC240TSA are displayed as curves with shading (mean ± measurement error). PT-AFM nano-DMA collected with an AC160TSA is displayed in previous figures. In (e), PT-AFM nano-DMA E' is calculated for both a non-adhesive contact with a spherical indenter (Hertz model, pink line), and an adhesive contact with a hyperboloid indenter (cyan line).

## CONCLUSIONS

In summary, we have shown that PT actuation of AFM cantilevers can be employed to perform nano-DMA experiments. Performing PT actuation with chirp signals provided a way to achieve viscoelastic quantification over a continuous and wide, five orders of magnitude, frequency range. Measuring sample viscoelasticity over such a broad frequency range in a continuous fashion is advantageous because such a capability abolishes the need to construct master curves by taking advantage of the TTS principle to shift data measured from a range of temperatures to a reference temperature within that range, as is currently the case with macro-DMA measurements. Avoiding the creation of master curves, while still measuring sample viscoelasticity over a broad and continuous frequency range, increases the ease of measuring the viscoelastic properties of samples that are sensitive to temperature changes or to being clamped and exposed to large deformations within a DMA apparatus. Additionally, for samples that are less sensitive to temperature, it may be possible to further expand the frequency range of PT-AFM nano-DMA measurements by performing measurements at multiple different temperatures (imposed by a temperature-controlled stage) and constructing master curves for the sample. Comparison of PT-AFM nano-DMA with macroscopic DMA data and PE-AFM nano-DMA suggests that the PT laser causes a local increase in sample temperature. Nevertheless, PT actuation does not involve magnetic fields that might potentially alter the sample, and avoids the presence of spurious peaks in the cantilever's spectrum typical of PE excitation. Therefore, in spite of the local temperature increase, elimination of disrupting stimuli (magnetic fields) and spurious resonances render PT-AFM nano-DMA more versatile for nano-DMA measurements. We note that although the measurement sensitivity is expected to increase when $k_c$ approaches $k_S$[41, 57], the correct application of PT-AFM nano-DMA requires that $k_c > |k^*|$. PT-AFM nano-DMA measurements are robust, and synergize well with measurements collected via other AFM techniques such as AM-FM AFM. The combination of PT-AFM nano-DMA and AM-FM AFM enhances the amount of information obtained about the sample by measuring sample viscoelasticity over a larger frequency range than either technique alone while also acquiring information about sample topography. To conclude, our novel PT-AFM nano-DMA technique serves as a useful tool to measure the nanoscale viscoelasticity of polymeric materials over a broad and continuous frequency range.

## EXPERIMENTAL METHODS

**Sample:** Styrene butadiene rubber (SBR) was chosen as test material for PT-AFM nano-DMA, because SBR is already well characterized by AFM experiments[17,19–22]. The SBR used was a random copolymer 36% styrene by weight, 57% 1,2-butadiene units in butadiene fraction, and 43% 1,4-butadiene units in butadiene fraction. The measured glass transition temperature was −13°C, determined by differential scanning calorimetry (DSC Q1000 TA Instruments). Samples were cut to size (see experiments below), stored at − 5°C, and let to equilibrate at room temperature prior measurements.

**Macroscale DMA Control:** Macroscale DMA experiments were performed on SBR samples in air, in order to serve as a control for nano-DMA measurements. A DMA Q800 (TA Instruments) in tension clamp configuration was used to perform DMA on an SBR sample 11.95 mm (L) x 4.70 mm (W) x 1.30 mm (H) in size. The sample was cooled to -50°C and stabilized for 20 mins, then clamped. Frequency sweeps were performed at 0.1 Hz, 0.3 Hz, 1 Hz, 3 Hz, and 10 Hz in 2°C steps from -50°C to 70°C. Before measurements at each step, the sample was allowed to equilibrate for 5 mins. Strain was maintained at 0.1% to ensure that DMA was performed in the SBR's linear viscoelastic regime. A 0.01 N preload force was used. Force track was set to 107%. The TA Instruments Rheology Advantage Data Analysis Software (TA Instruments) was used to calculate the shift factors from the tanδ curve in order to generate DMA master curves. The master curves for E', E'', and tanδ were calculated for a reference temperature $T_0$ = 40°C. For the TTS principle, the calculated shift factors can be used to shift the master curves for different temperatures[1,3].

**AFM:** AFM was performed on small SBR samples, roughly 2 mm (L) x 2 mm (W) x 3 µm (H) in size. All AFM experiments were performed in air at room temperature with a Cypher ES AFM (Oxford Instruments Asylum Research). AC160TSA-R3 (Olympus, nominal spring constant 26 N m$^{-1}$, resonance frequency 300 kHz, tip radius 7 nm, tetrahedral tip shape), AC240TSA-R3 (Olympus, nominal spring constant 2 N m$^{-1}$, resonance frequency 70 kHz, tip radius 7 nm tetrahedral tip shape), and biosphere™-NT_B2000_v0010 (Nanotools, nominal spring constant 40 N m$^{-1}$, resonance frequency 330 kHz, tip radius 2 µm, spherical tip shape, used in the Supporting Information SI4) cantilevers were used. Cantilever calibration was performed by indenting a hard substrate and using the thermal noise method[49]. Data acquisition and cantilever calibration were performed using the Oxford Instruments Asylum Research V16 software based in Igor Pro.

For **AFM nano-DMA**, sinusoidal excitations were applied to the cantilever. Sinusoidal cantilever excitation was achieved using exponential (also called logarithmic) chirped signals to provide mechanical measurements over a continuous and wide frequency range (see Supporting Information SI3).

The broad frequency range of nano-DMA measurements was achieved by combining data obtained performing measurements with three different chirp regimes with different start and end frequencies, in order to optimize the sampling frequency for each measurement range and avoid software crashes caused by processing large amounts of data. The frequency range of each chirp was: (i) 0.1 Hz - 10.1 Hz, (ii) 1 Hz - 1,001 Hz, and (iii) 200 Hz - 20,200 Hz. The sweep time of each chirp was: (i) 180 s, (ii) 30 s, and (iii) 1 s. For (i,ii), data were recorded in the time domain with a sampling rate of (i) 1,000 Hz, and (ii) 10,000 Hz. Data were then transformed in the frequency domain using Fast Fourier Transform (FFT). Finally, signals were normalized by the driving signal and smoothed using the Savitzky-Golay filter. For (iii), the lock-in amplifier accessible in the Cypher ES AFM used was employed. The lock-in sampling rate was set to 5,000 Hz, and the low-pass filter time constant to 100 Hz. Chirps (i)-(iii) were performed both out of contact with the sample as a reference measurement, and in contact with the sample for a sample measurement. At least three reference and sample measurements were performed, and k', k'', E', E'', and tanδ were calculated using the average amplitude and phase of the reference and sample measurements. The resulting signals from (i)-(iii) were then combined in order to quantify sample viscoelasticity over the entire frequency range.

For sample chirps, the sample was indented with an approach/withdraw velocity of 1 µm s$^{-1}$ and a trigger point of 100 nN. Approach and withdraw time were 5 s. Withdraw distance was set to 3 µm. After the 5 s approach, an exponential chirp was applied to the sample. Contact points were identified using the force-indentation variation (FIV) method[55], and corrected manually in the event that the FIV method did not accurately identify the contact point. The same procedure was followed for both PT- and PE- AFM nano-DMA measurements.

For **PT-AFM nano-DMA** experiments, PT actuation was achieved by focusing an excitation laser (EL, blueDrive™, Oxford Instruments Asylum Research, 405 nm) at the base of the AFM cantilever (see Supporting Information SI4 for details on positioning the PT-EL). The DC voltage applied to the PT-EL photodiode was set to 4 V, and the AC voltage to 1 V for AC160 and 0.1 V for AC240 cantilevers to obtain similar reference amplitudes between cantilevers. Reference measurements were collected approximately 500 µm above the sample surface. See Supporting Information SI4 for details on the effect of reference measurement distance above the sample.

For **PE-AFM nano-DMA** experiments, PE actuation was achieved by placing an external piezo actuator (PL088.31 PICMA® Chip Actuators, Physik Instrumente Ltd, 10 mm x 10 mm x 2 mm) underneath the sample, as in other studies[18–20,26]. The piezo actuator was secured to a metallic disk with double-sided tape. The SBR sample and a thin glass slide for reference measurements were placed on another metallic disk. The disk containing the sample was fixed with double-sided tape on top of the piezo actuator. The piezo actuator was then connected to the AFM electronics to control piezo actuator motion, with an applied AC voltage of 1 V. Typical resulting cantilever motion can be found in Supporting Information SI1.

**Data analysis and visualization** for macroscale DMA, as well as PT- and PE-AFM nano-DMA measurements were performed in Python with home-built codes.

**AM-FM AFM imaging** was performed on dry SBR samples as a control for PT-AFM nano-DMA measurements. AM-FM AFM creates high-resolution maps of sample viscoelasticity by oscillating a cantilever at two eigenmodes[15]. Sample tanδ is measured at the lower frequency (f1) by the first eigenmode [15,51,52]. Sample E' is measured by both modes, but the higher frequency mode (f2) contributes most to the contact stiffness, and hence the value of E'[14]. Therefore, E' corresponds to sample properties at the higher frequency (f2)[14,15,56]. The Hertz contact model is applied in order to measure sample E', but not tanδ[14,15,51,52,56]. In these experiments, PT actuation was used to achieve AM-FM AFM's bimodal cantilever excitation.

For SBR AM-FM AFM, both AC240 (nominal kc,1 ~2 N/m, kc,2 ~ 50 N/m, f1 ~ 70 kHz, f2 ~ 400 kHz) and AC160 (nominal kc,1 ~26 N/m, kc,2 ~ 364 N/m, f1 ~ 300 kHz, f2 ~ 1.5 MHz) were used. Cantilevers were excited using the same PT EL used for PT-AFM nano-DMA measurements. Multiple spots on the SBR surface were scanned, resulting in a total of 25 AM-FM AFM images taken with the AC240 and 13 with the AC160. After the experiment, all AM-FM AFM images of SBR topography were flattened using Asylum Research software version 16.10.208 in Igor Pro software version 6.38B01, in order to remove any variations in sample topography that were not due to SBR features. This flattening was done by hand, in order to avoid introducing flattening artifacts. The processed files were then analyzed by a custom script in MATLAB R2019b that calculated tanδ and E' via the formulas in refs.[14,15,51,52,56], then compared AM-FM AFM measurements to macroscale DMA control values at the relevant frequency.

## Author Contributions

A.R.P designed the PT-AFM nano-DMA technique. A.R.P. performed the theoretical calculations and determined the signal processing for AFM nano-DMA. A.R.P. and C.A. performed the AFM experiments and data analysis. A.R.P. and N.H. performed the macroscale DMA experiments and data analysis. R.W. implemented the code used to perform AFM nano-DMA to Asylum Research's V16 software. J.S., R.W., Y.T., and R.P. provided helpful insight and design suggestions for AFM nano-DMA and data analysis. A.R.P., C.A., and S.C. wrote the paper with input from all other authors. S.C. supervised the work. All authors extensively discussed the results.

## Funding Sources

ARP acknowledges funding from a UK Engineering and Physical Sciences Research Council (EPSRC) graduate scholarship, and additional support from the Sidney Perry foundation, the Blanceflor Boncompagni Ludovisi née Built foundation, the Sapienza University of Rome for the Borsa di Perfezionamento all'Estero, and the Angelo Della Riccia foundation.

## ACKNOWLEDGMENTS

The authors thank Stephen Duncan for helpful insights on signal processing, Ileana Andrea Bonilla-Brunner for insightful discussions on AFM and samples, Antoine Jérusalem and Charles Clifford for extensive discussions, and the Asylum Research Oxford Instruments team, especially Aleksander Labuda and Jonathan Moffatt, for technical support.

## ABBREVIATIONS

AC, alternating current; AFM, atomic force microscopy; AM-FM, amplitude modulation-frequency modulation; CR, contact resonance; DC, direct current; DMA, dynamic mechanical analysis; FIV, force-indentation variation; FTT, Fast Fourier Transform; JKR, Johnson-Kendall-Roberts; PE, piezoelectric; PT, photothermal; SBR, styrene-butadiene rubber; TTS, time-temperature superposition;

## Supporting Information

# Nanoscale rheology: Dynamic Mechanical Analysis over a broad and continuous frequency range using Photothermal Actuation Atomic Force Microscopy

Alba R. Piacenti[a], Casey Adam[a,b], Nicholas Hawkins[b], Ryan Wagner[c], Jacob Seifert[a], Yukinori Taniguchi[d], Roger Proksch[e], Sonia Contera[a]*

[a]Clarendon Laboratory, Department of Physics, University of Oxford, OX1 3PU Oxford, United Kingdom

[b]Department of Engineering Science, University of Oxford, OX1 3PJ Oxford, United Kingdom

[c]School of Mechanical Engineering, Purdue University, West Lafayette, Indiana, 47907, United States

[d]Asylum Research, Oxford Instruments KK, Tokyo 103-0006, Japan

[f]Asylum Research-An Oxford Instruments Company, Santa Barbara, California 93117, United States

* Corresponding author. E-mail: sonia.antoranzcontera@physics.ox.ac.uk

**This document contains the following sections:**

### SI1: SPURIOUS PEAKS IN PE, BUT NOT PT-AFM NANO-DMA MEASUREMENTS

Often in PE excitation, spurious resonances arising from the piezo and the piezo/AFM system add noise to AFM measurements[1–3]. Figure S1 shows typical cantilever amplitude and phase signals obtained with PT and PE-AFM nano-DMA, respectively direct and indirect ways to excite the AFM cantilever. At frequencies above ~1 kHz, PE amplitude and phase exhibit multiple local minima and maxima, which are spurious peaks that limit the frequency range over which PE-AFM nano-DMA can be performed. These spurious peaks are not present for PT-AFM nano-DMA.

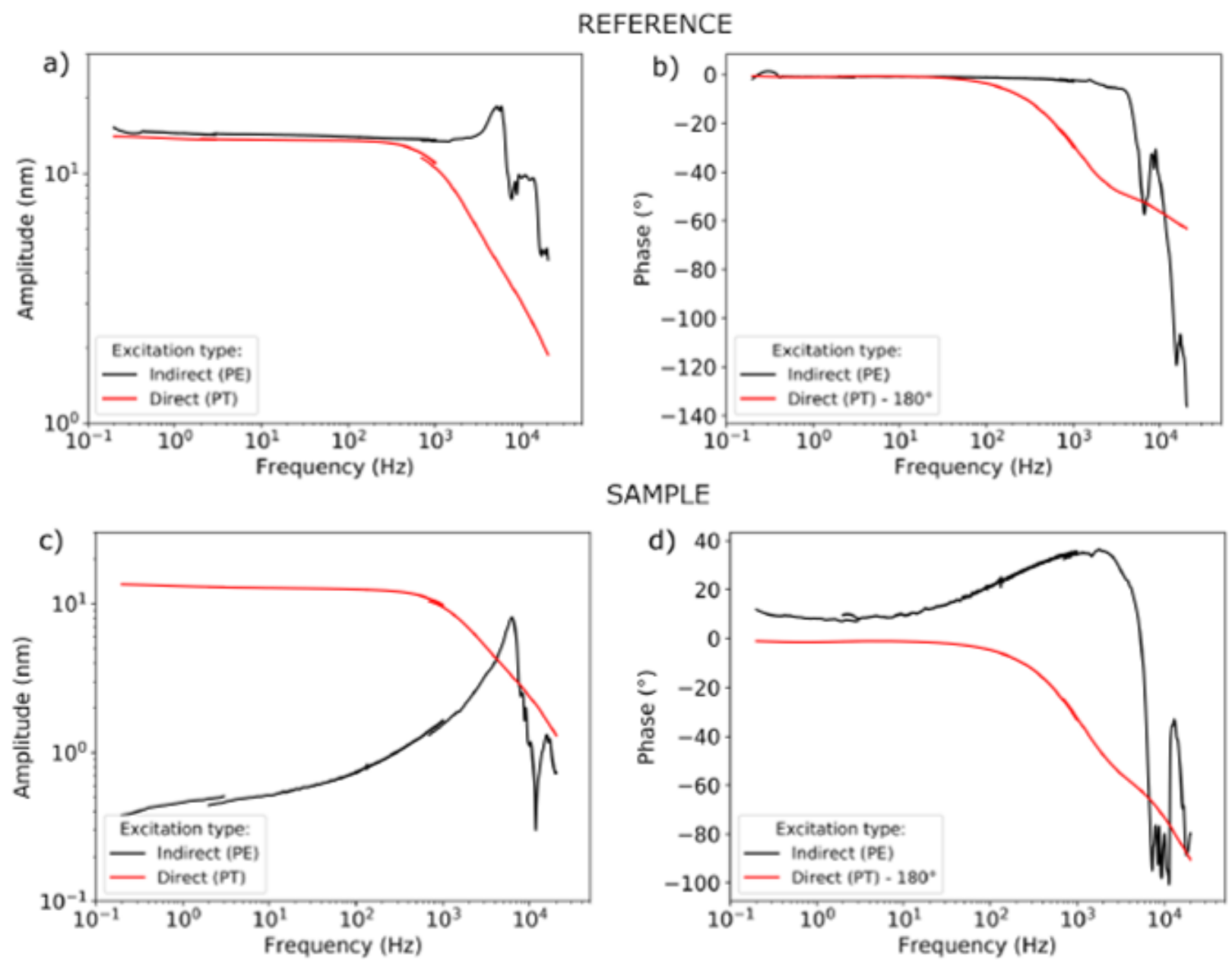

Figure S1. Spurious peaks in PE, but not PT measurements. Cantilever amplitude is shown in (a) for reference measurements and (c) for sample (SBR) measurements. Similarly, cantilever phase is shown in (b) and (d) for reference and sample (SBR) measurements, respectively. PT measurements (red lines) are smooth for all frequencies in the measured range while PE measurements (black lines) exhibit multiple peaks. These spurious peaks hinder mechanical characterization above ~1 kHz using PE actuation.

## SI2: CONTACT MODELS USED FOR E* CALCULATION

### SIMPLIFICATION OF EQUATIONS FOR HYPERBOLOID INDENTER GEOMETRY

For indenters with hyperboloid shapes, Sun et al.[4] proposed the following model to describe the AFM cantilever-sample contact when adhesion forces are present:

$$d = \frac{aA}{2R_I}\left[\frac{\pi}{2} + arcsin\left(\frac{\left(a/A\right)^2 - 1}{\left(a/A\right)^2 + 1}\right)\right] - \sqrt{\frac{2\pi w a}{\tilde{E}}} \qquad (S1)$$

$$F = \frac{2\tilde{E}A}{2R_I}\left[aA + \frac{a^2 - A^2}{2}\left(\frac{\pi}{2} + arcsin\left(\frac{\left(a/A\right)^2 - 1}{\left(a/A\right)^2 + 1}\right)\right)\right] - \sqrt{8\pi a^3 w\tilde{E}} \qquad (S2)$$

Where d is the indentation, F the force, $a$ the contact radius between the indenter (I) and the sample (S), $\tilde{E}$ the reduced Young's modulus, $w$ is the energy of adhesion per unit contact area, and A = $R_I$ cotα, with α being the indenter semi-vertical angle and $R_I$ the indenter (AFM tip) radius.

For the AC160 cantilevers used in this study, $a_1 \approx 200$ nm (details on obtaining $a_1$ are provided in SI6, which discusses calculating the error in E' and E'') and A ≈ 22 nm (for α = 17.5°, $R_I$ = 7nm). Therefore, $a \gg A$ and the following approximation can be used:

$$arcsin\left(\frac{\left(a/A\right)^2 - 1}{\left(a/A\right)^2 + 1}\right) \approx arcsin(1) = \frac{\pi}{2} \qquad (S3)$$

Using this approximation, Eqs. (S1) and (S2) simplify to:

$$d = \frac{aA\pi}{2R_I} - \sqrt{\frac{2\pi wa}{\tilde{E}}} \qquad (S4)$$

$$F = \frac{2\tilde{E}A}{2R_I}\left[aA + \frac{a^2 - A^2}{2}\pi\right] - \sqrt{8\pi a^3 w\tilde{E}} \qquad (S5)$$

### "*TWO-POINTS METHOD*" FOR SHERICAL, HYPERBOLOID, AND CONICAL INDENTERS

In this section, equations used for the application of the "*two-points method*"[4] using specific points in force-indentation curves are provided for indenters with different geometries. As explained in the main text (and Figure 4), these specific points are the point of zero load "0", the point around which dynamic oscillations occur "1", and the point of zero deformation "2".

#### Spherical indenter geometry

For indenters with a JKR spherical contact, the contact radius can be calculated as follows[5]:

$$a^3 = \frac{\tilde{R}}{K}\left(F + 3\pi\tilde{R}w + \sqrt{6\pi w\tilde{R}F + \left(3\pi w\tilde{R}\right)^2}\right) \qquad (S6)$$

With $\tilde{R} = \frac{R_I R_S}{R_I + R_S} \approx R_I$ given that the size of the sample (S) is much larger than that of the indenter (I).

Equation S6 can then be used to calculate the contact area at different points along the force indentation curve in order to apply the "*two-points method"* for the spherical JKR contact as follows[4,5]:

Indentation can be written in terms of $a$ and $a_0$[4,5]:

$$d = \frac{a^2}{\tilde{R}}\left[1 - \frac{2}{3}\left(\frac{a_0}{a}\right)^{\frac{3}{2}}\right] \qquad (S7)$$

$$a_0^3 = \frac{6w\pi\tilde{R}^2}{K} \qquad (S8)$$

By using point "2" and point "0", $K$ can be calculated as[4]:

$$K = -\frac{F_2}{2}\frac{3}{\sqrt{d_0\tilde{R}}} \qquad (S9)$$

And $a_0$ can be calculated from Eq.(S7) as follows[4]:

$$a_0 = \sqrt{3d_0\tilde{R}} \qquad (S10)$$

$w$ can be then calculated by combining Eq.(S9) and (S10) with Eq.(S8). Finally, $a$ can be calculated with Eq.(S6).

#### Hyperboloid and conical indenter geometry

Calculated in point "0", Eq.(S4) becomes:

$$\sqrt{\frac{2a_0 w\pi}{\tilde{E}}} = \frac{a_0 A}{2R_I}\pi - d_0 \qquad (S11)$$

From which:

$$\frac{w}{\tilde{E}} = \frac{\left(\frac{a_0 A}{2R_I}\pi - d_0\right)^2}{2a_0\pi} \qquad (S12)$$

Eq.(S5) calculated in point "0" becomes:

$$F_0 = 0 = \frac{A}{2R_I}\left(a_0 A + \frac{a_0^2 - A^2}{2}\pi\right) - a_0\sqrt{\frac{2\pi a_0 w}{\tilde{E}}} \qquad (S13)$$

Applying equation (S11) to equation (S13) gives the following relation:

$$\frac{A}{2R_I}\left(a_0 A + \frac{a_0^2 - A^2}{2}\pi\right) - a_0\left(\frac{a_0 A}{2R_I}\pi - d_0\right) = 0 \qquad (S14)$$

From which $a_0$ can be extracted. The SymPy Python library was used in the data analysis to extract $a_0$. Once $a_0$ is known, $\frac{w}{\tilde{E}}$ can be calculated from Eq.(S12).

Calculated in point "1", Eq.(S7) becomes:

$$d_1 = \frac{a_1 A}{2R_I}\pi - \sqrt{2a_1\pi}\sqrt{\frac{w}{\tilde{E}}} \qquad (S15)$$

From which $a_1$ can be obtained. The SymPy Python library was used in the data analysis to extract $a_1$.

Eq.(S8) calculated in point "1" becomes:

$$F_1 = 2\tilde{E}\left[\frac{A}{2R_I}\left[a_1 A + \frac{a_1^2 - A^2}{2}\pi\right]\right] - \left[a_1\sqrt{2a_1\pi}\sqrt{\frac{w}{\tilde{E}}}\right] \qquad (S16)$$

With algebraic rearrangement, Eq.(S16) becomes:

$$\tilde{E} = \frac{F_1}{2\left[\frac{A}{2R_I}\left[a_1 A + \frac{a_1^2 - A^2}{2}\pi\right]\right] - \left[a_1\sqrt{2a_1\pi}\sqrt{\frac{w}{\tilde{E}}}\right]} \qquad (S17)$$

$w$ can be then calculated from Eq.(S12)

Similarly, the same procedure used for the hyperboloid indenter can be applied for conical indenters using Eqs.(5,6) with $d_{NA} = \frac{\pi}{2} a \cot\alpha$ and $F_{NA} = \frac{\pi}{2} a^2 \tilde{E} \cot\alpha$ [6].

## SI3: ANALYSIS OF CHIRP SIGNALS TO EXCITE THE AFM CANTILEVER

Most AFM techniques that measure nanoscale viscoelasticity use single frequency sinusoidal signals to excite the cantilever, thereby quantifying sample viscoelasticity at specific, discrete frequencies[7–18]. However, in order to measure viscoelasticity over a continuous range of frequencies, the single-frequency excitation sinusoid can be replaced with chirp signals[19–22]. Chirp signals have already been employed for PE-AFM rheology to provide continuous and wide frequency range spectra up to 10 kHz[21,22] or 30 kHz[19].

Fig. S2 shows the amplitude and phase of a typical PT-AFM nano-DMA reference measurement obtained by exciting the cantilever with a linear and an exponential, also called logarithmic, chirp signal. Fringes are present at frequencies less than 100 Hz in the amplitude and phase of cantilevers excited with linear chirp signals. These fringes are not present when exponential chirps are used. For this reason, exponential chirps should be used to excite the cantilever during PT-AFM nano-DMA experiments.

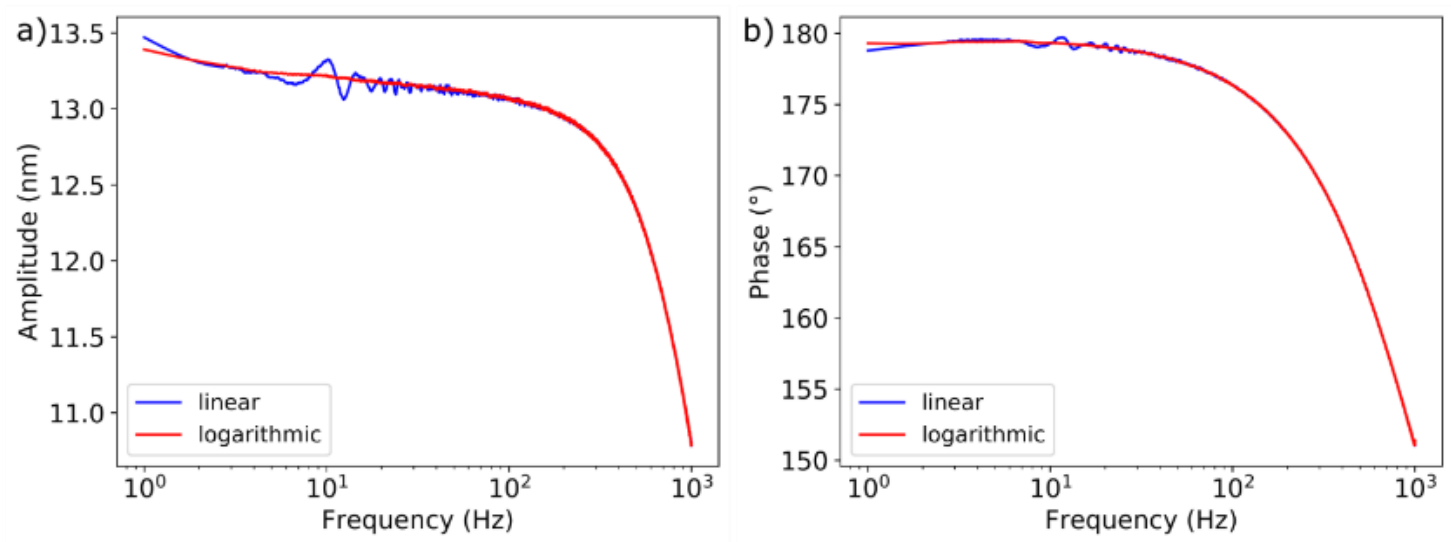

Figure S2. Comparison of cantilever oscillations excited by linear or exponential, also called logarithmic, chirps. The amplitude (a) and phase (b) of cantilever oscillations driven by linear chirp signals (blue line) and logarithmic chirp signals (red line) are shown. Fringes appear in cantilever amplitude and phase at frequencies less than 100 Hz for cantilevers driven by a linear chirp.

## SI4: ANALYSIS OF EXPERIMENTAL FACTORS THAT COULD AFFECT CANTILEVER AND SAMPLE RESPONSES DURING PT-AFM NANO-DMA MEASUREMENTS

A detailed analysis of factors that could influence the cantilever and sample responses during a PT-AFM nano-DMA experiment is performed in this section. These factors include (A) PT-EL position on the cantilever, (B) PT-EL spot drift, (C) PT-EL power, (D) PT-EL drive amplitude, (E) effect of the trigger point, the extent of the stimulus applied to the sample, (F) cantilever approach/withdraw velocity, (G) cantilever selection and the relationship between sample vs. cantilever stiffness, and (H) the distance from the sample surface at which the reference measurement is collected. Finally (I), this analysis is used to recommend a procedure that should be applied to each new sample in order to optimize PT-AFM nano-DMA control parameters and ensure reliable PT-AFM nano-DMA quantification of sample viscoelasticity.

### A. Effect of PT Laser Position

PT excitation laser (EL) position on the cantilever alters the amplitude of coated[23–27] and uncoated[24,27,28] cantilever oscillations, and is important in correctly quantifying material stiffness[25]. Therefore, PT-EL positioning could potentially alter PT-AFM nano-DMA measurement values. To investigate whether PT-AFM nano-DMA measurements are robust to different PT-EL positions, PT-AFM nano-DMA was performed on the SBR with a variety of different PT-EL positions, shown in Figure S3. For convenience in the following discussion, each PT-EL position is defined by two coordinates. The first coordinate represents position along the cantilever's length, x/L, where x/L = 1 is at the end of the cantilever and x/L = 0 is at the base. The second coordinate represents position along the cantilever's width, y/W, where y/W = 0 represents the center of the cantilever, and y/W = 1/3 or −1/3 represents the side of the cantilever.

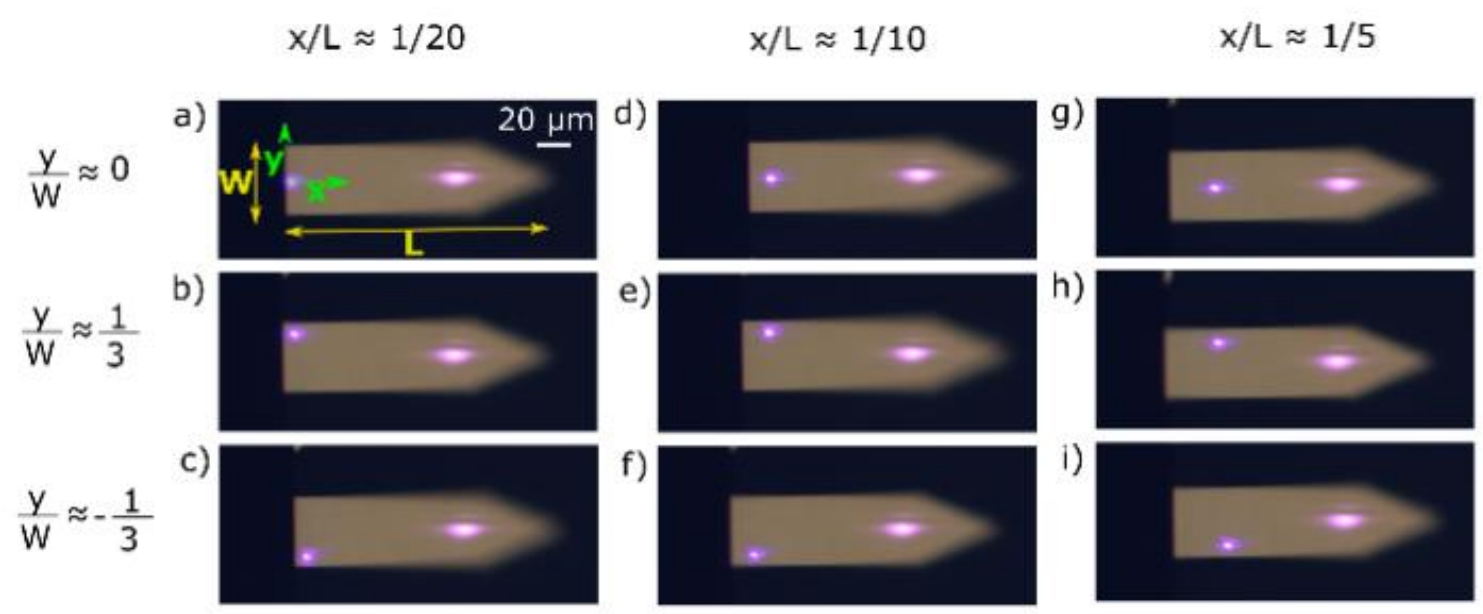

Figure S3. Different PT-EL positions for PT-AFM nano-DMA measurements. The x coordinate, written above the images, represents position along the cantilever's length (L), where x/L = 1 is at the end of the cantilever and x/L = 0 is at the base of the cantilever. Similarly, the y coordinate, written to the left of the images, represents position along the cantilever's width (W), where y/W = ± 1/3 is at the edge of the cantilever, and y/W = 0 is in the center of the cantilever.

Figure S4 shows the effect of PT-EL position on the amplitude and phase signals for both reference and sample measurements. The value of y/W had a negligible effect on cantilever amplitude and phase compared to x/L. At low frequencies, reference phase measurements were robust to PT-EL position. At high frequencies, reference phase decreased as x/L→ 1. Sample phase measurements followed a similar trend to reference measurements, but the effect was less pronounced. For both reference and sample measurements regardless of frequency, cantilever amplitude increased as x/L → 1.

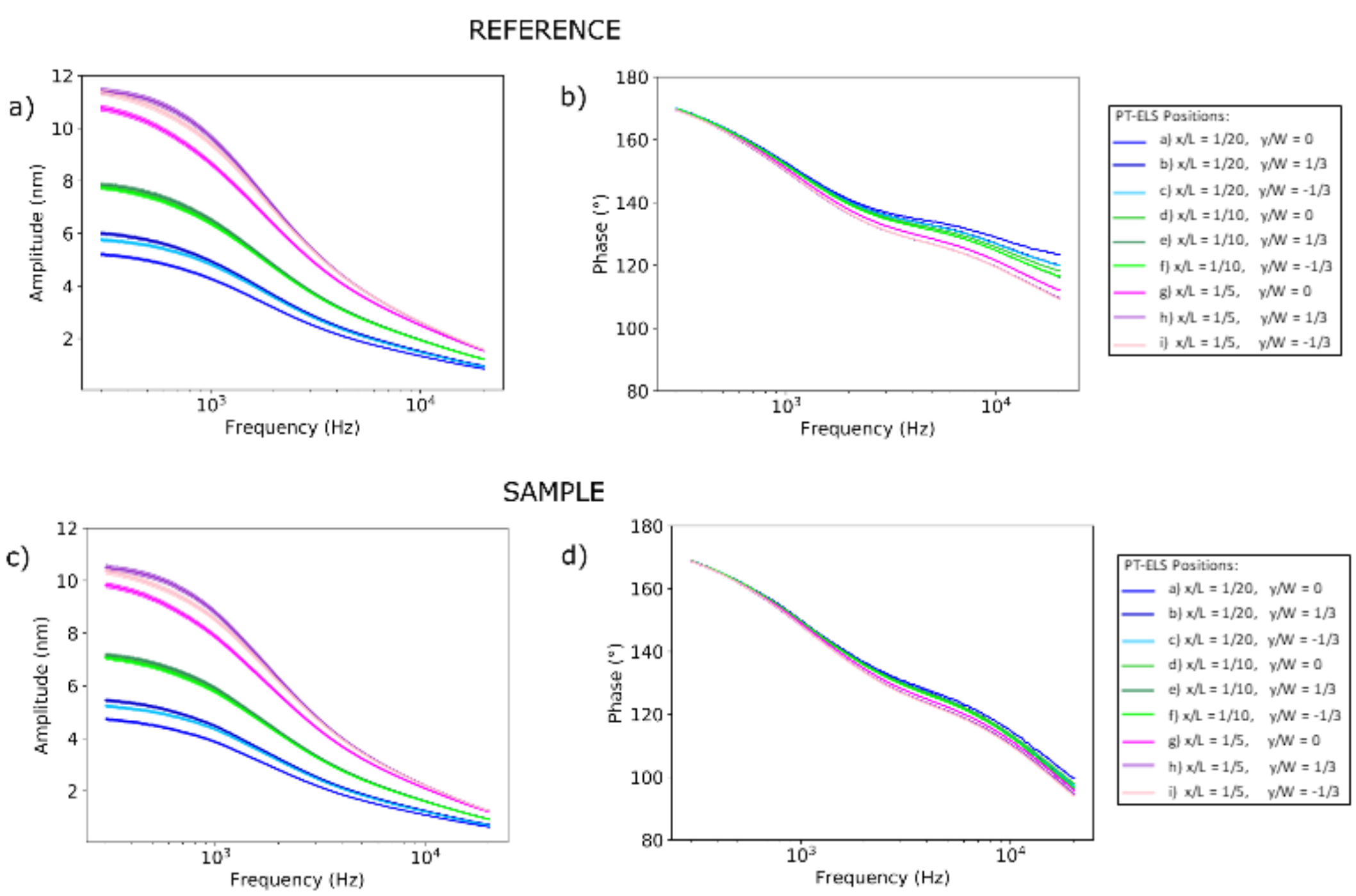

Figure S4. Effect of PT-EL position on cantilever amplitude and phase. Reference amplitude and phase are shown in (a) and (b), respectively. Similarly, sample amplitude and phase are shown in (c) and (d), respectively. PT-EL positions displayed in the legend correspond to those in figure S3.

Figure S5 shows the effect of PT-EL position on PT-AFM nano-DMA quantification of SBR tanδ. As with amplitude and phase measurements (Figure S4), y/W did not alter tanδ values significantly. As x/L -> 1, the measured tanδ decreased compared to measurements obtained with smaller x/L values. However, also shown in Figure S5, PT-AFM nano-DMA quantification of sample tanδ agreed fairly well with the macroscale DMA measurements in spite of tanδ shifts due to PT-EL position changes. Therefore, PT-AFM nano-DMA measurements are robust to PT-EL position, in spite of the differences in cantilever amplitude and phase. However, the combined results of Figures S4 and S5 suggest that PT-EL positioning at $x/L \sim 1/10$ is ideal because cantilever amplitude is highest in this position and the measured tanδ has the best agreement with the macroscale DMA control. While y/W had little effect on cantilever motion, $x/L \sim 1/10$ and $y/W \sim 0$ is preferable, because this position has already been shown to provide reliable off-resonance mechanical measurements[25]. Therefore, PT-EL positioning at $x/L \sim 1/10$ and $y/W \sim 0$ should be used during PT-AFM nano-DMA measurements.

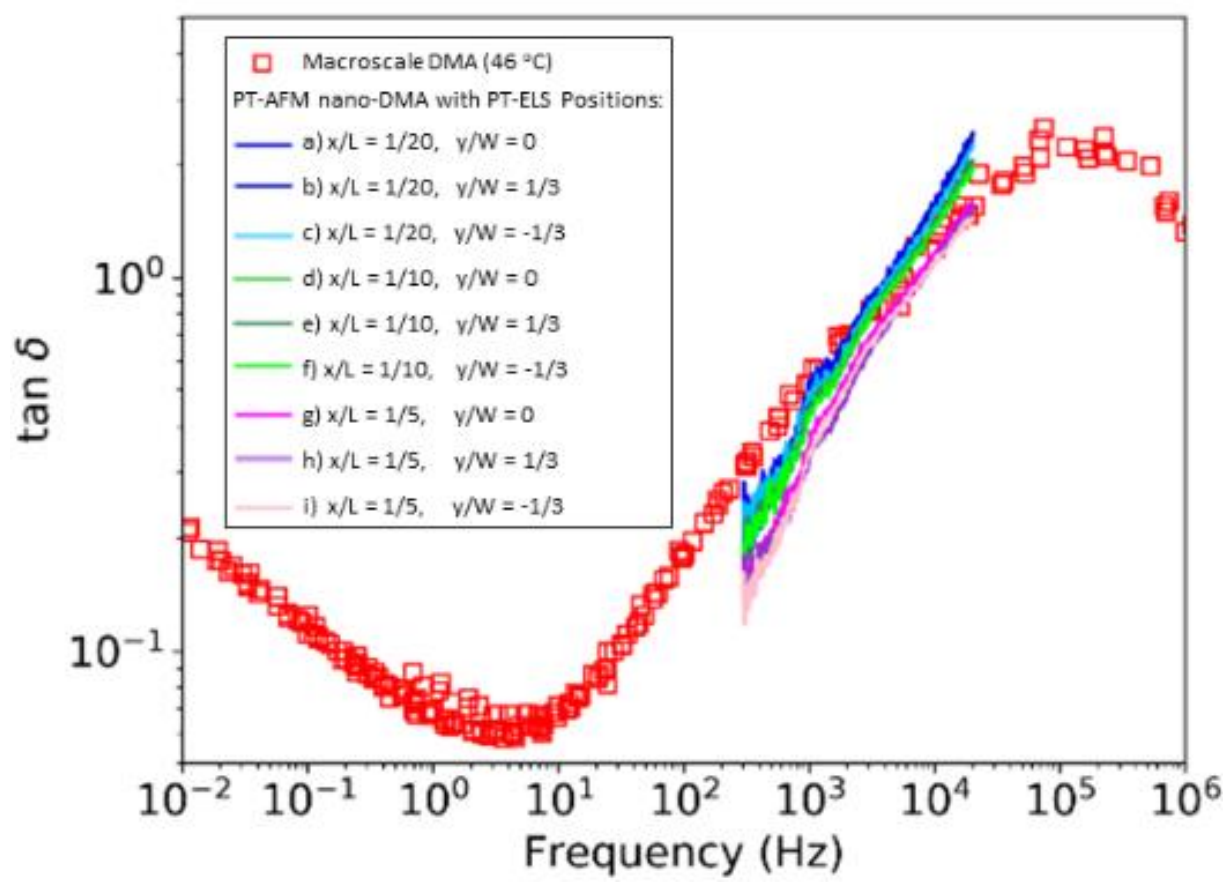

Figure S5. Effect of PT-EL position on loss tangent quantification. SBR loss tangent (tanδ) measurements collected with different PT-EL positions are shown as lines. PT-EL positions displayed in the legend correspond to those in Figure S3. Macroscopic DMA control values are shown as unfilled red squares.

## B. Effect of Laser Spot Drift

During an AFM experiment, it is possible that the PT-EL and detection laser positions on the cantilever could drift. Therefore, it is necessary to determine whether PT-AFM nano-DMA measurements are robust to laser spot drift. After zeroing the photodiode, the following procedure was adopted to determine the effects of laser spot drift on PT-AFM nano-DMA measurements. First, a reference measurement was performed 500 μm above the sample surface. Second, the tip was lowered into contact with the sample and a sample measurement was performed. Third, the cantilever was raised to 600 μm above the surface, then lowered to 550 μm, then again lowered to 500 μm above the sample surface. In this third step, the cantilever was not immediately moved to 500 μm above the sample in order to avoid measuring the effects of cantilever backlash, defined as temporary changes in cantilever position due to motor direction reversal, and thereby only measure effects due to laser spot drift. The entire procedure was repeated five times without zeroing the photodiode.

Figures S6 and S7 show the effect of PT-EL spot drift on the amplitude and phase of reference and sample measurements, as well as the relative measurement errors. For a signal x, the relative error is $|x - M| / M$, where M is the mean of all x. Relative errors were less than 1% and 0.25% for amplitude and phase. Gibbs phenomena occur at the edges of the measured frequency range, resulting in larger relative errors. This larger error at the edges of the frequency range is therefore expected and can be discounted.

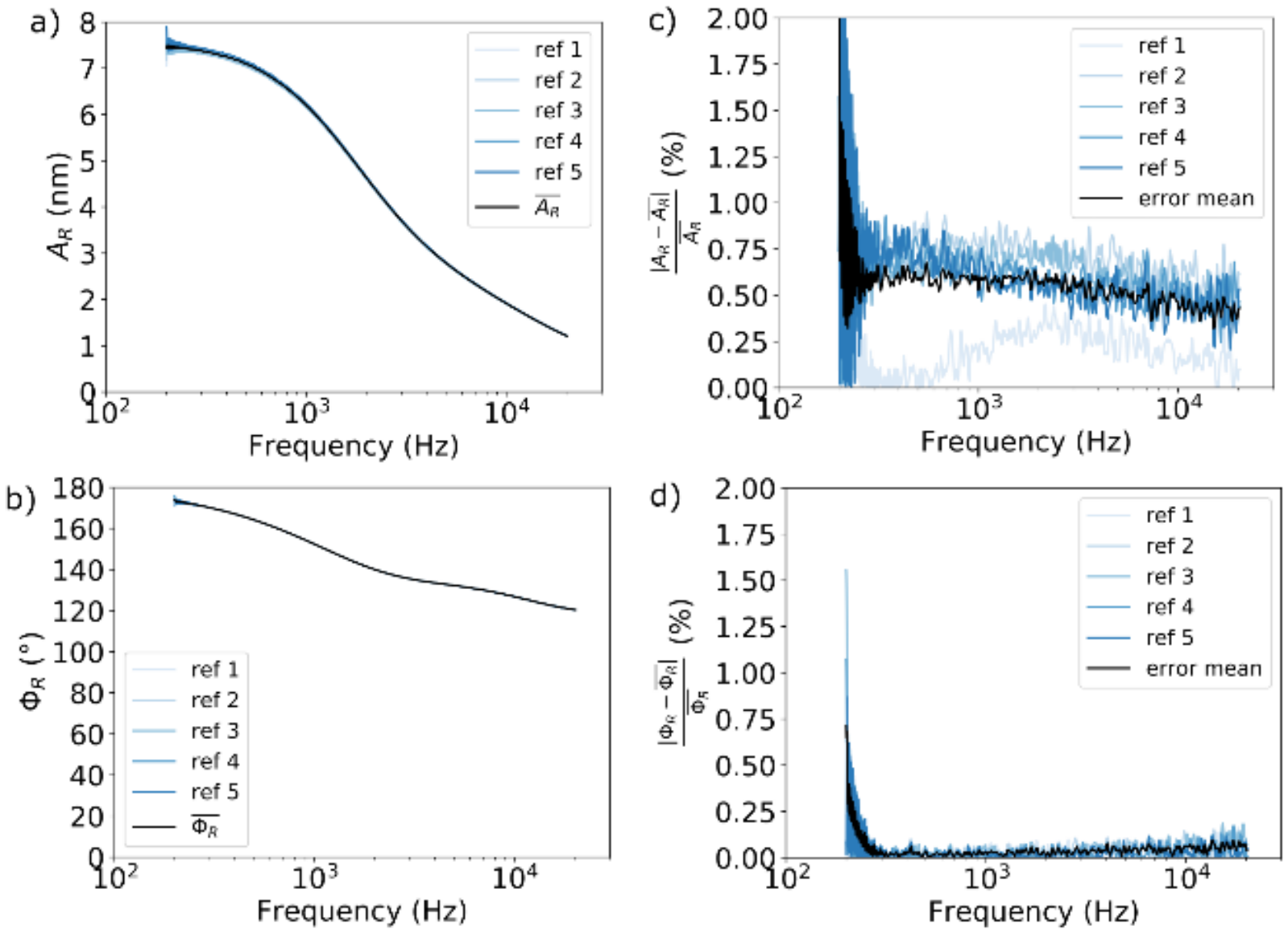

Figure S6. Effect of PT-EL spot drift on reference measurement amplitude (A) and phase ($\phi$). Reference (subscript R) A and $\phi$ measured throughout the course of laser spot drift are shown in (a) and (b), respectively. The relative error between A and $\phi$ measurements subjected to different drifts are shown in (c) and (d), respectively.

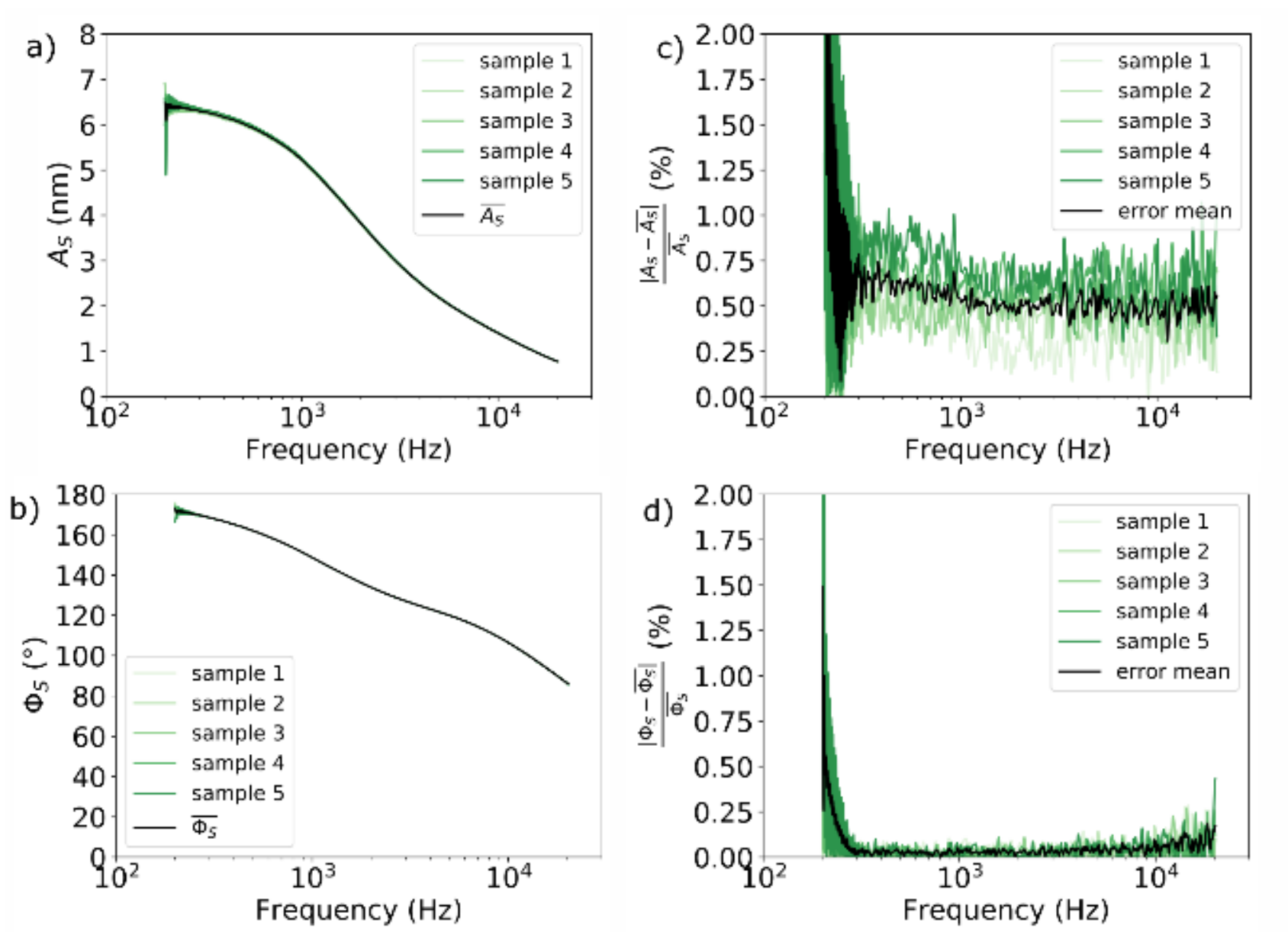

Figure S7. Effect of PT-EL spot drift on sample measurement amplitude (A) and phase ($\phi$). Sample (subscript S) A and $\phi$ measured throughout the course of laser spot drift are shown in (a) and (b), respectively. The relative error between A and $\phi$ measurements subjected to different drifts are shown in (c) and (d), respectively.

### C. Effect of PT Laser Power

The power of the PT-EL affects how much the cantilever bends, and therefore could potentially alter PT-AFM nano-DMA measurements. In order to determine whether PT-EL power affected quantification of sample viscoelasticity,

PT laser power was modulated via filters built into the Cypher AFM. Figure S8 shows cantilever amplitude and SBR tanδ measured by PT-AFM nano-DMA performed at 100%, 30%, and 10% of the maximum PT-EL power. For all three powers, a drive amplitude of 1V was used. Reference and sample amplitudes decreased with decreasing PT-EL power. As PT-EL power decreased, measured tanδ values shifted slightly to lower frequencies, and exhibited increased noise. However, regardless of PT-EL power, the overall shape of tanδ vs. frequency remained similar to the macroscopic DMA control. The shift to lower frequencies for lower PT-EL powers is likely due to the fact that lower PT-EL powers heat the sample less than higher PT-EL powers. The increased tanδ noise at lower PT-EL powers is likely due to the lower cantilever amplitudes, and hence worse signal-to-noise ratios at lower PT-EL powers. For measurements at high frequencies collected at 10% PT-EL power, cantilever amplitude decreased below the nm scale. This amplitude decrease is the most likely reason why PT-AFM nano-DMA tanδ results deviate from the DMA control at high frequencies with 10% PT-EL power. Together, these results suggest that, as long as cantilever amplitude is sufficiently high (nm or higher), PT-AFM nano-DMA is robust to laser power. If PT-EL power is too low, discontinuities and additional peaks appear in the measured tanδ. On the other side, it should be also considered that if PT-EL is too high there is the risk of introducing unwanted non-linear viscoelastic effects due to too large cantilever oscillations (and consequent sample deformations). Therefore, it is important to perform PT-AFM nano-DMA at different PT-EL powers for each new sample, in order to determine the optimal PT-EL power at which to perform the measurements.

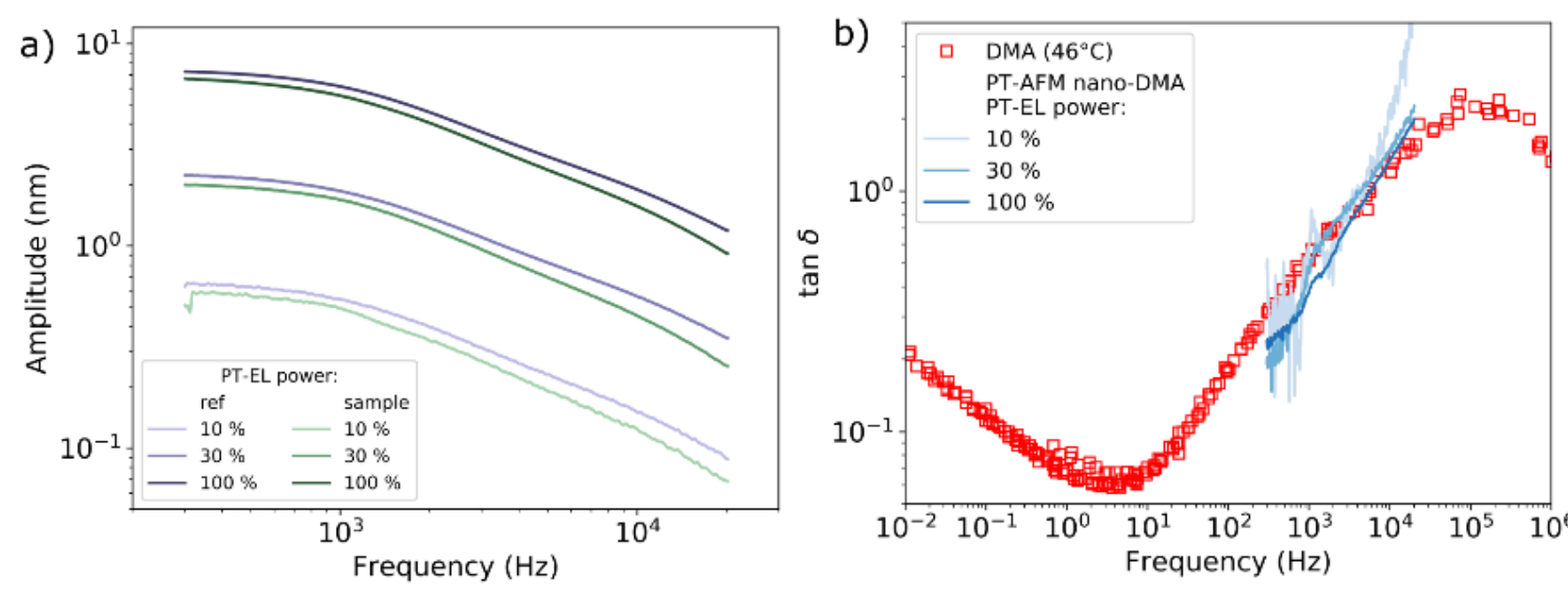

Figure S8. Effect of PT-EL power on PT-AFM nano-DMA measurements. Cantilever amplitude at different PT-EL powers is shown in (a). SBR tanδ measured by PT-AFM nano-DMA at different powers (lines) and macroscopic DMA (unfilled squares) are shown in (b).

**D. Effect of Drive Amplitude**

As described in the previous section, ensuring that cantilever amplitude is at least in the nm range is critical in correctly quantifying sample viscoelasticity via PT-AFM nano-DMA. Users control cantilever amplitude by changing PT-EL power, as in the previous section, or more commonly by changing the drive amplitude applied to the PT-EL. In order to further investigate the effects of cantilever amplitude on PT-AFM nano DMA measurements, SBR measurements were performed with a drive amplitude of: 0.1V, 0.25V, 0.50V, 1.00V, and 1.5V. At each drive amplitude, two reference and two sample measurements were performed.

Figure S9 shows the relationship between drive amplitude and the resulting reference and sample amplitudes during PT-AFM nano-DMA. Measurement amplitudes were linearly proportional to the drive amplitude (Figure S9(c,d)).

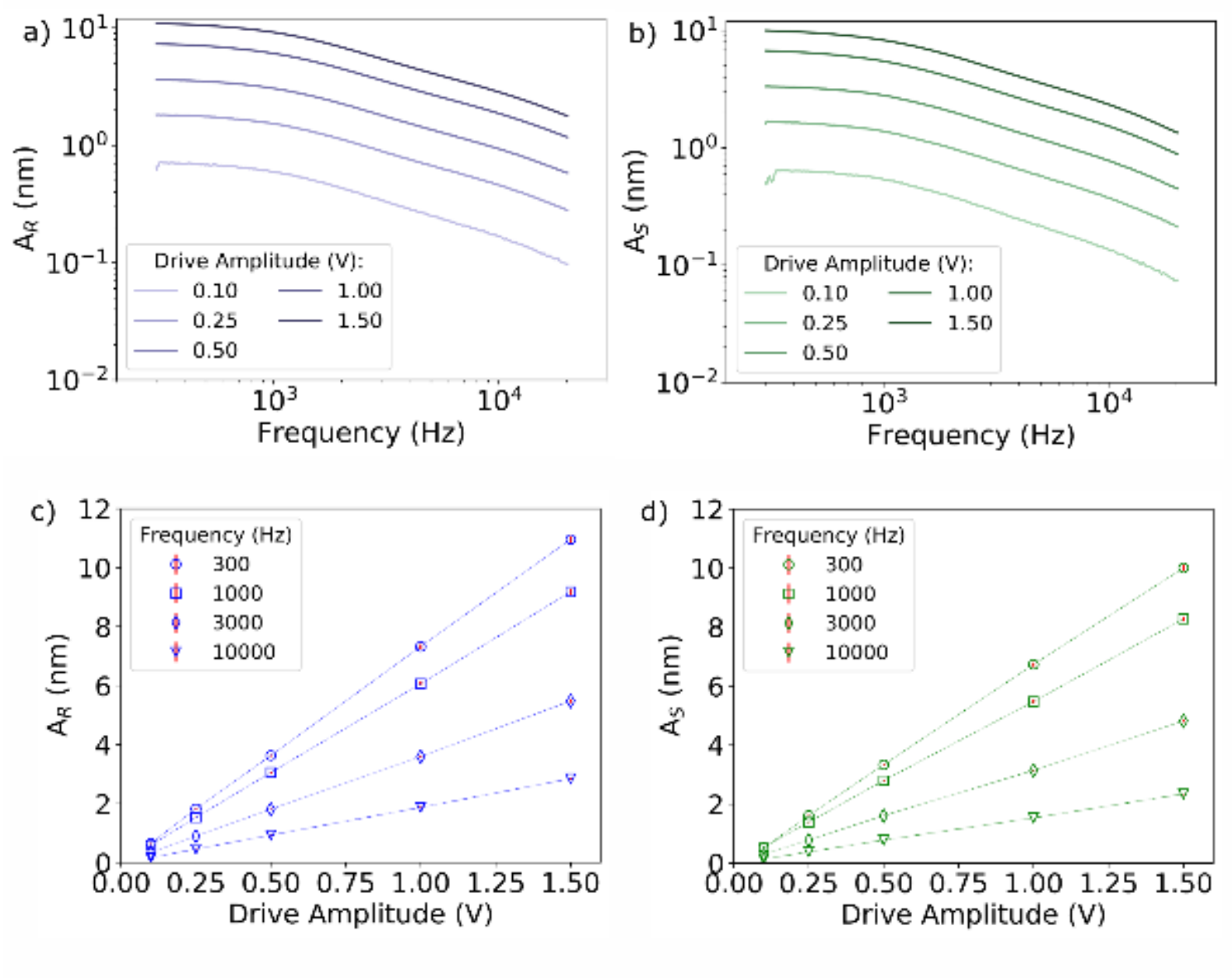

Figure S9. Effect of drive amplitude on cantilever amplitude. The amplitude (A) of cantilever oscillations during reference (R) and sample (S) measurements are shown in (a) and (b), respectively. The relationship between A and drive amplitude at specific frequencies is shown in (c) and (d) for R and S measurements, respectively.

Since cantilever amplitude increased with drive amplitude, it is possible that increasing the drive amplitude too much will result in sample deformations outside of the SBR's linear viscoelastic regime. If PT-AFM nano-DMA measurements occur in the linear viscoelastic regime of a sample, a linear relationship should exist between the reference and sample amplitudes because the response of the sample to the stimulus is linear. That is, the amplitude of the oscillatory force applied to the sample ($\propto A_R$) will cause a linear increase in the oscillatory deformation of the sample ($A_S$). It is important to note that the slope of this relationship at different frequencies might vary because a sample's viscoelastic response is frequency dependent. As shown in figure S10, measurements were performed in the SBR's linear viscoelastic regime for all drive amplitudes. The slope of the line at different frequencies varied, indicating differences in SBR viscoelasticity at different frequencies.

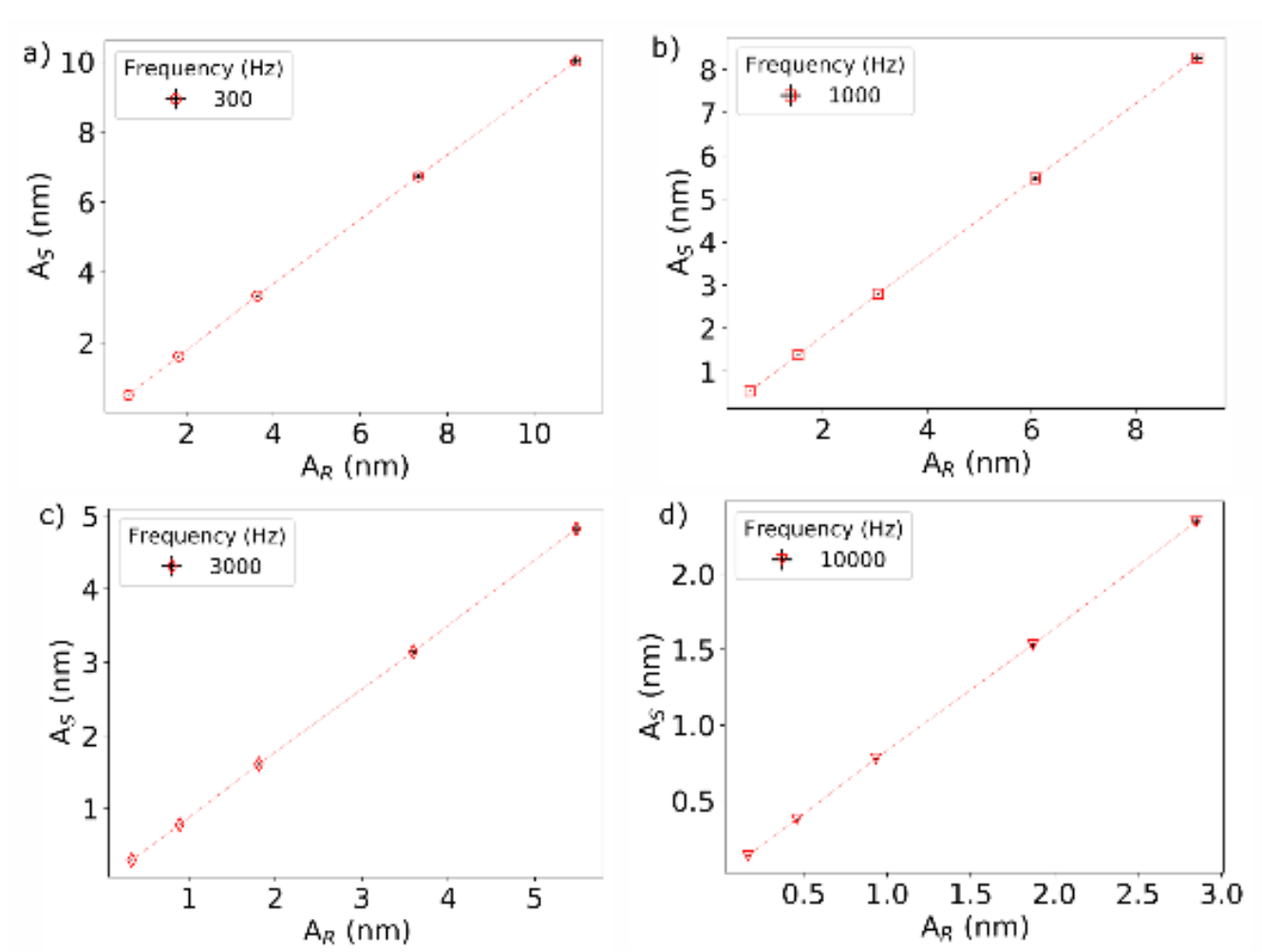

Figure S10. Linear viscoelasticity of the SBR. If PT-AFM nano-DMA is performed in the linear viscoelastic regime of a sample, a linear relationship should exist between cantilever amplitudes (A) during the sample (S) and reference (R) measurements. This line is shown in (a)-(d) at frequencies of 300 Hz, 1,000 Hz, 3,000 Hz, and 10,000 Hz.

SBR tanδ measured at each drive amplitude were similar, as shown in Figure S11, validating that the measurements were performed within the linear viscoelastic regime of the sample. However, as drive amplitude decreased, the resulting tanδ curves grew increasingly noisy. Therefore, for any given sample, if tanδ measurements appear noisy or bumpy, it may be a sign that the measurement amplitude is too low, and that the drive amplitude should therefore be increased.

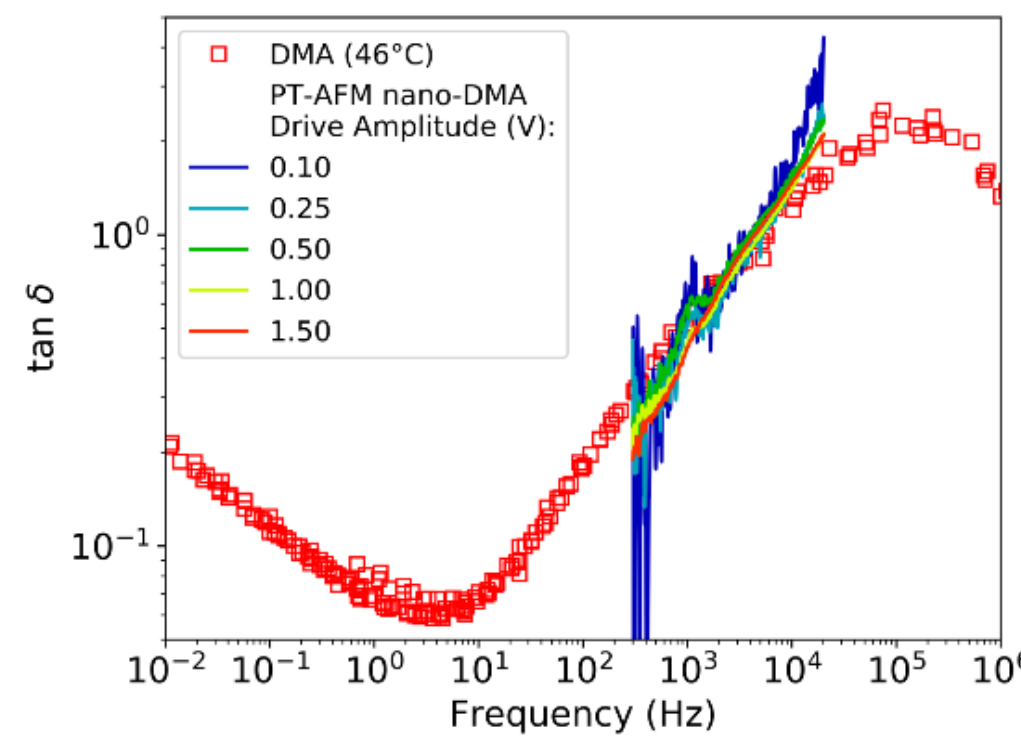

Figure S11. Effect of drive amplitude on loss tangent (tanδ) measurements. SBR tanδ measured by PT-AFM nano-DMA at different drive amplitudes (lines) and macroscopic DMA (unfilled squares) are shown.

**E. Effect of Force Trigger Point**

As mentioned in the previous section, a sample's response to stimulus depends on whether sample deformations caused by the stimulus fall within the sample's linear viscoelastic regime. Initial sample indentation, before the chirp oscillation is applied, can potentially affect PT-AFM nano-DMA results if the applied force deforms the sample beyond the linear viscoelastic regime. In order to determine the effects of the initial applied force on SBR PT-AFM nano-DMA, sample measurements were performed at different force trigger points of: 10 nN, 25 nN, 50 nN, 75 nN, 100 nN, 250 nN, 500 nN, and 1,000 nN, which led to increasing indentation depths. All sample measurements were compared to the same reference measurement.

As shown in figure S12 (a,b), k‘ and k'' increased with increasing initial applied force. This observation is unsurprising, since stiffness is an extensive quantity that depends on the geometry of the contact which is affected by the indentation depth. Additionally, k′ at high frequencies exhibits a sudden decrease for large trigger points for this sample and cantilever combination. This change in k' causes a sudden increase, similar to an asymptote, in tanδ (Figure S12(c)). The appearance of the asymptote-like changes in k' and tanδ suggest that higher trigger points resulted in sample deformations outside the SBR's linear viscoelastic regime. Therefore, it is important to ensure that the trigger point selected for PT-AFM nano-DMA measurements is within the sample's linear viscoelastic regime.

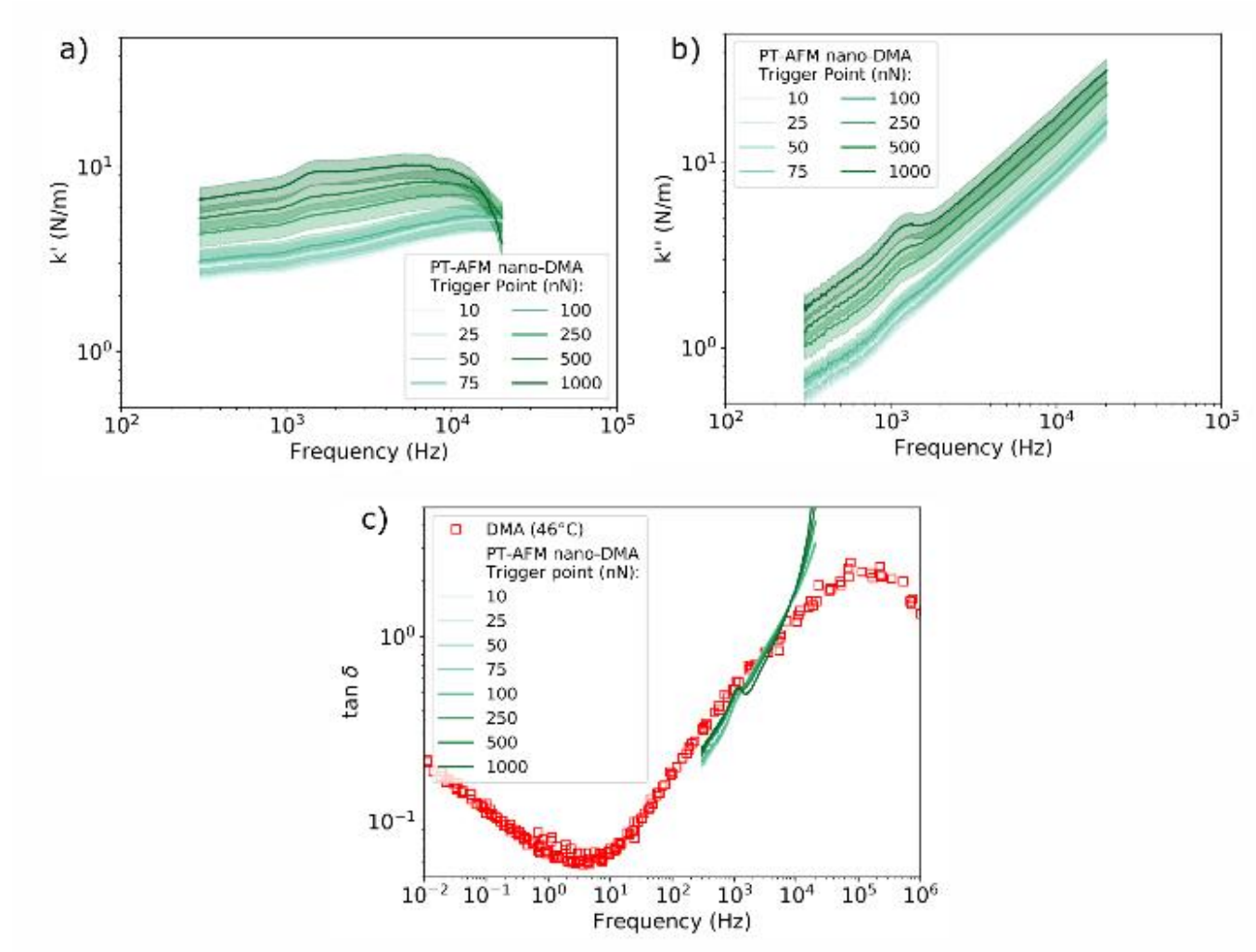

Figure S12. Effect of initial applied force on PT-AFM nano-DMA measurements. SBR k' and k'' measured at different initial applied forces are shown in (a) and (b), respectively. The resulting tanδ is shown in (c). PT-AFM nano-DMA measurements are shown as lines, while macroscopic DMA control measurements are shown as unfilled squares.

### F. Effect of Approach and Withdraw Velocity

The velocity at which a sample is indented can affect a sample's viscoelastic response. For slower velocities, polymers within the sample have more time to rearrange in response to the tip. For faster velocities, polymers have less time to rearrange. In order to determine whether cantilever approach/withdraw velocity altered PT-AFM nano-DMA measurements, the SBR sample was indented at velocities of: 0.1 μm $s^{-1}$, 1.0 μm $s^{-1}$, 10.0 μm $s^{-1}$, 50.0 μm $s^{-1}$, and 100 μm $s^{-1}$. Approach time was set to 5 s for all velocities except 0.1 μm $s^{-1}$, where the approach time was reset to 35 s in order to give the cantilever enough time to contact the sample before applying the oscillations. As shown in figure S13, approach/withdraw velocity did not alter PT-AFM nano-DMA measurements in this experiment.

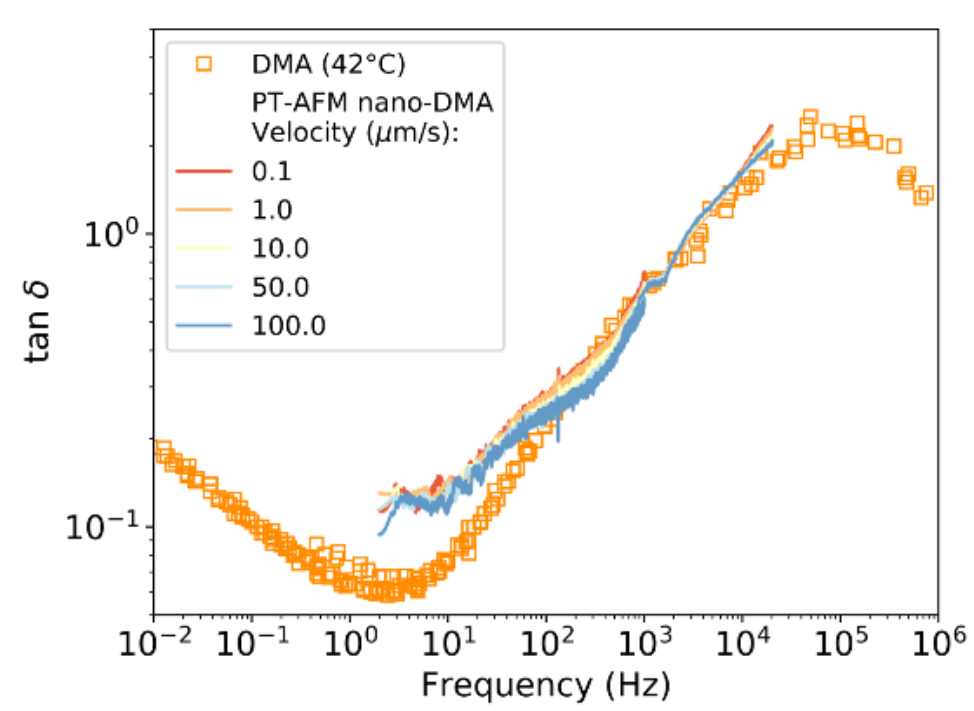

Figure S13. Effect of approach/withdraw velocity on PT-AFM nano-DMA measurements. SBR tanδ measured by PT-AFM nano-DMA with different cantilever approach/withdraw velocities are shown as lines. Macroscopic DMA control values are shown as unfilled squares.

### G. Effect of Contact Stiffness

The ratio of dynamic stiffness (k*) and cantilever stiffness ($k_c$) can affect PT-AFM nano-DMA measurements. If |k*| >> $k_c$, the sample is not indented enough to measure. If |k*| << $k_c$, the measurement is insensitive to changes in

sample modulus. Additionally, differences in optical lever calibration between the freely vibrating cantilever (reference) amplitude and the in-contact vibrating cantilever (sample) amplitude must be small for the equations used to calculate PT-AFM nano-DMA k' and k'' (equations 1 and 2) to apply[25]. To minimize these differences in optical lever calibration, the cantilever used should satisfy $k_C > |k^*|$ [25].

To examine the effect of cantilever selection and contact stiffness on SBR measurements, PT-AFM nano-DMA was performed on the SBR using different cantilevers. Three cantilevers were tested. The first cantilever was an AC160TSA-R3 (Olympus, nominal tip radius 7 nm, tetrahedral tip shape, calibrated spring constant 16.93 N $m^{-1}$, resonance frequency 267 kHz), the second cantilever was an AC240TSA-R3 (Olympus, nominal tip radius 7 nm, tetrahedral tip shape, calibrated spring constant 1.13 N $m^{-1}$, resonance frequency 55 kHz,), and the third cantilever was a biosphere™-NT_B2000_v0010 (Nanotools, nominal tip radius 2 µm, spherical tip shape, calibrated spring constant 38.92 N $m{-1}$, resonance frequency 284 kHz.). Cantilever stiffness was calibrated by indenting a hard substrate, then using the thermal noise method[30]. The drive amplitude was varied for each cantilever so that reference measurements collected by all three cantilevers had approximately the same reference amplitude.

Three reference and three sample measurements were collected for each cantilever and averaged. The resulting k', k'', and tanδ values are shown in figure S14. Biosphere k' and k'' are higher than those of the AC240 and AC160 cantilevers, due to the altered contact geometry of a 2 µm tip (Biosphere) vs. a 7 nm tip (AC160 and AC240). Asymptote-like deviations from control DMA measurements appear for Biosphere and AC240 measurements when |k*| approaches the stiffness of each cantilever. The same does not occur for AC160 cantilevers because |k*| was always lower than AC160 cantilever stiffness. Therefore, these results demonstrate that, as long as $k_C > |k^*|$ over the entire frequency range, PT-AFM nano-DMA can accurately quantify sample viscoelasticity. For each sample, it is therefore important to compare |k*| to cantilever stiffness in order to ensure this condition holds.

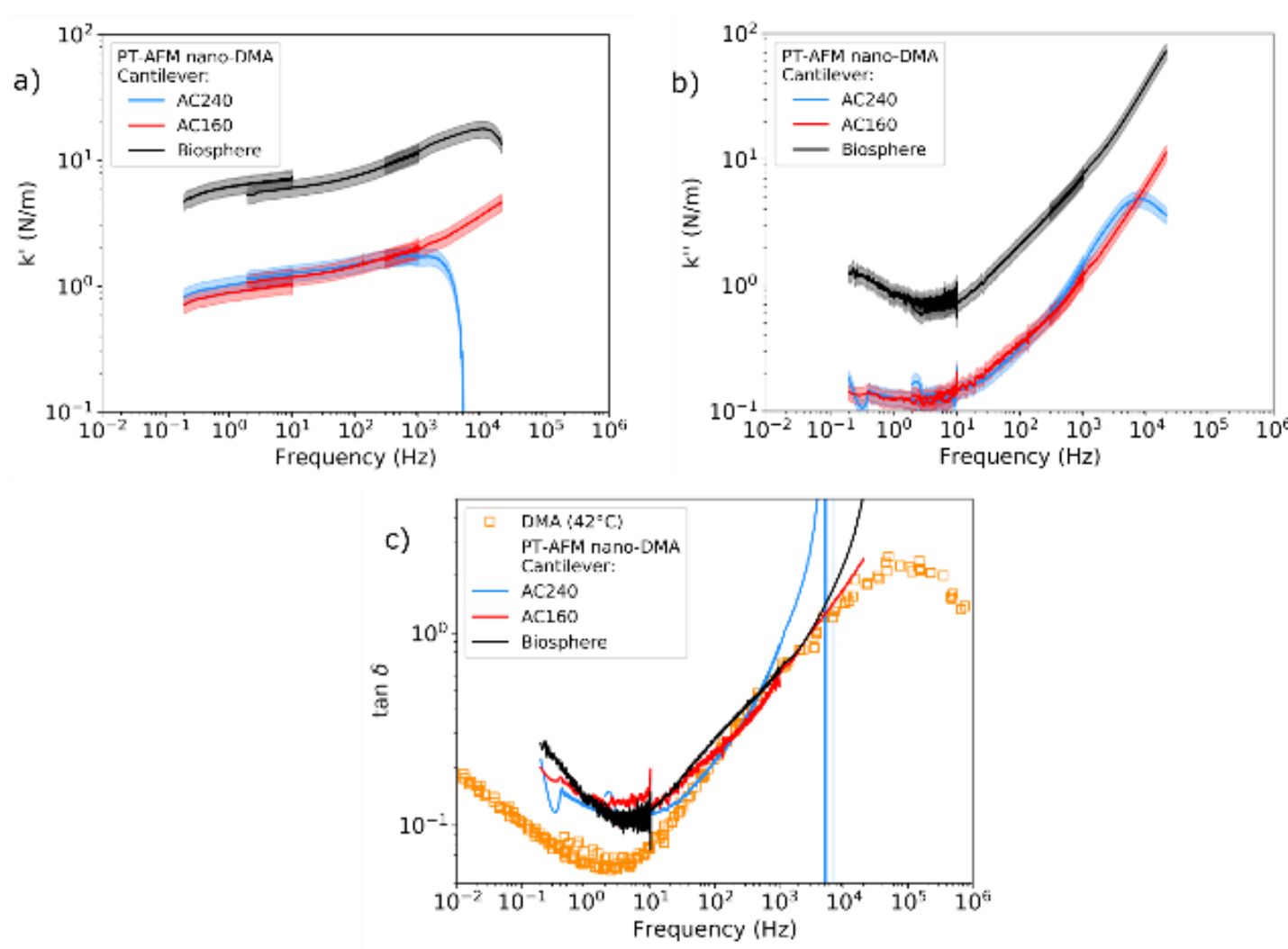

Figure S14. Effect of cantilever selection and contact stiffness on PT-AFM nano-DMA measurements. PT-AFM nano-DMA measurements performed with different cantilevers, and therefore different contact stiffness, are shown as lines. The resulting k', k'', and tanδ are shown in (a-c), respectively. Macroscale DMA control measurements are shown as unfilled squares.

## H. Effect of Reference Measurement Height

Surface effects, such as the moisture layer over dry samples or squeeze film damping, might influence reference measurements if the reference is collected too close to a sample's surface. In order to determine the effect of reference measurement height, three reference measurements were collected and averaged at a distance of 500 µm, 400 µm, 300 µm, 200 µm, 100 µm, 50 µm, 3 µm, and 1 µm from the sample surface. After collecting four reference measurements at each height, four sample measurements were performed and averaged. The average sample measurement was then compared to the average reference measurement at each height.

Figure S15 shows the amplitude and phase of each reference measurement. All reference measurement amplitudes, regardless of reference height, were within the calculated error (see section SI4 B) of one another. Therefore, reference height did not affect the reference amplitude. However, as shown in Figure S15(b), references collected 1 µm and 3 µm from the sample surface had an approximate phase difference of 0.2° from the other reference measurements. This observation suggests that the SBR sample may have influenced the 1 µm and 3 µm reference measurements.

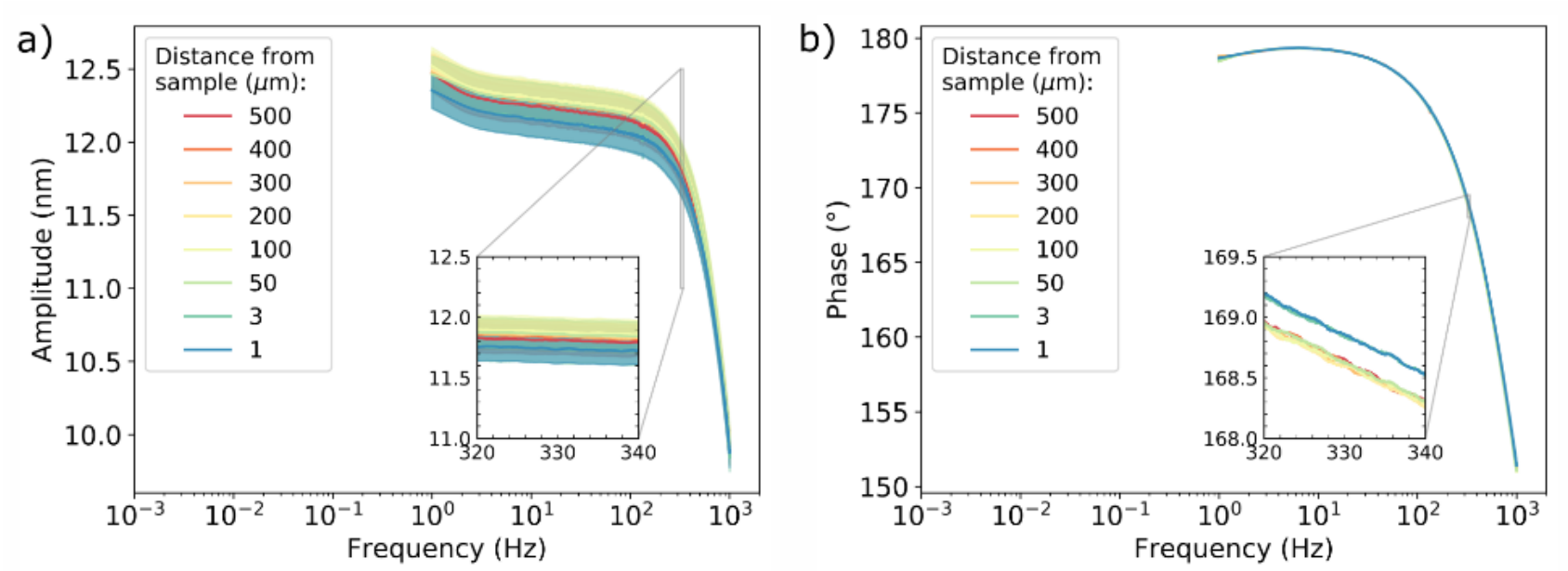

Figure S15. Effect of reference measurement height on cantilever amplitude and phase. Cantilever amplitude and phase are shown in (a) and (b), respectively, for measurements collected at different heights above the sample surface.

Figure S16 shows the effect of reference height on PT-AFM nano-DMA quantification of SBR viscoelasticity. Quantification of SBR tanδ agrees with the DMA control measurements for all reference heights, save for references collected 1 µm and 3 µm from the sample surface. This observation supports the notion that the closer reference measurements were subject to surface effects. Therefore, PT-AFM nano-DMA measurements of sample viscoelasticity are robust over a wide range of reference heights as long as the reference is not subject to surface layer effects.

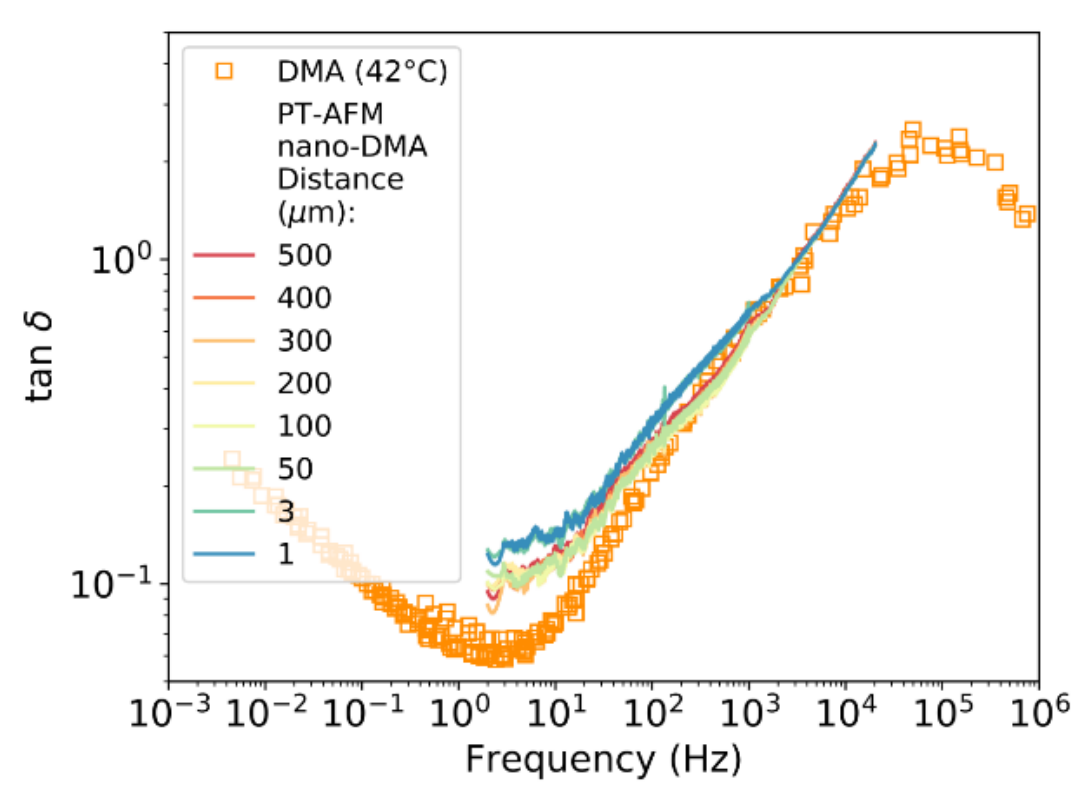

Figure S16. Effect of reference measurement height on the measured tanδ. PT-AFM nano-DMA measurements with references at different heights are shown as lines. Macroscale DMA control measurements are shown as unfilled squares.

## I. Summary

In summary, PT-AFM nano-DMA is robust to a variety of factors that can influence the cantilever and sample response. Based on the analyses in this section, when performing PT-AFM nano-DMA on dry samples, it is important to ensure that the PT-EL is positioned in the center of the cantilever, and roughly 1/10 of the way down the length of the cantilever. It is also important to ensure that (i) cantilever amplitude is large enough to be above the detection noise floor but not so large that sample deformations are beyond the sample's linear viscoelastic regime, (ii) that the initial applied stimulus is in the sample's linear viscoelastic regime, (iii) that |k*| never exceeds cantilever stiffness, and (iv) that the reference is far enough from the sample surface to avoid any surface effects. Incorrect selection of any of the parameters in (i)-(iv) results in incorrect measurement of sample viscoelasticity, including noisy data and asymptotes that appear in the measured k', k'', and tanδ curves. Therefore, when PT-AFM nano-DMA is first performed on a sample, it is ideal to account for these factors by collecting measurements with: different trigger points, different PT-EL powers and drive amplitudes, and different reference heights to find the optimal combination to measure the given sample's viscoelasticity. From these combinations, ideal PT-AFM nano-DMA parameters for the given sample can be selected by minimizing the amount of noise in the measurement (for example tanδ) curves and excluding parameters that result in asymptote-like changes in the curves.

## SI5: k* AND tanδ ERROR CALCULATION

The uncertainty $dy$ of a quantity $y$ that is a product or ratio of $N$ different $x_i$ that each have an associated uncertainty $dx_i$ ($i$ = 1, … , $N \in \mathbb{N}$ ) is calculated as follows[29]:

$$\frac{dy}{|y|} = \sqrt{\sum_{i=1}^{N}\left(\frac{dx_i}{x_i}\right)^2} \tag{S18}$$

For the PT-AFM nano-DMA components of k* calculated using Eqs. (1,2), the error could arise from uncertainty in cantilever stiffness $k_C$, and the cantilever amplitude and phase. Since the relative error was much larger for amplitude measurements than for phase measurements (see section SI4 B), only amplitude effects were used to calculate the uncertainty in PT-AFM nano-DMA measurements due to PT-EL spot drift. Therefore, the error in k' and k'' calculated with Eq. (1,2) can be calculated as follows:

$$\frac{dk^j}{|k^j|} = \sqrt{\left(\frac{dk_c}{k_c}\right)^2 + \left(\frac{d\overline{A}}{\overline{A}}\right)^2} \tag{S19}$$

Here, $k^j$ represents k' and k''. The value of $dk_c \,/\, k_c$ for cantilever's stiffness evaluated with the thermal noise method[30] can be estimated as 0.15[31]. Since the relative error in the amplitude is 0.01, the overall error in PT-AFM nano-DMA amplitude measurements due to laser spot drift is 1.4%, calculated as follows:

$$\frac{d\overline{A}}{\overline{A}} = \sqrt{\left(\frac{dA_R}{A_R}\right)^2 + \left(\frac{dA_S}{A_S}\right)^2} = \sqrt{0.01^2 + 0.01^2} = 0.014 \tag{S20}$$

Here, R and S indicate a reference or sample measurement, respectively. Therefore, the amplitude error (1.4%) is much smaller compared to the uncertainty in cantilever stiffness (15%).

Recalling that tanδ = k''/k', the error in tanδ can be calculated as follows.

$$\frac{dtan\delta}{tan\delta} = \frac{\partial\left(\frac{\overline{A}sin\Phi}{\overline{A}cos\Phi - 1}\right)}{\partial\overline{A}}d\overline{A} = \frac{tan\delta}{\overline{A}cos\Phi - 1}\frac{d\overline{A}}{\overline{A}} \tag{S21}$$

## SI6: E* ERROR CALCULATION

The error in AFM nano-DMA quantification of sample E' and E'', can be calculated from Eq.(4) as with Eq.(S.18):

$$\frac{dE^j}{E^j} = \sqrt{\left(\frac{dk}{k^j}\right)^2 + \left(\frac{da_1}{a_1}\right)^2} \quad (S22)$$

where j indicates the storage (') and the loss ('') moduli.

The uncertainty in $a_1(da_1)$ can be written as the contribution of 2 factors: $da^{curve}$, the error in the quantities used to calculate $a_1$ (i.e. variation of $F_1$ and $d_1$), and $da^{tip}$, the error in the indenter's geometrical parameters (e.g. tip radius, α):

$$da = da^{curve} + da^{tip} \quad (S23)$$

$da_1^{curve}$ can be calculated as the standard deviation of $a_1$ estimated over several curves.

$da_1^{tip}$ can be estimated by calculating how much $a_1$ changes by altering the tip's geometrical parameters compared to the nominal values.

In Figure S17, $a_1$ values calculated using different tip geometrical parameters for an AC160 cantilever are plotted with their standard deviation obtained from several curves ($da_1^{curve}$).

Figure S17(a) shows $a_1$ calculated for an AC160 with the JKR model with indenter radii close to the nominal tip radius (R=7nm). For a blunted cantilever with R=14 nm $da_1^{tip} = |a_1^{R=7nm} - a_1^{R=14}|$. From Figure S17(a), it can be seen that $da_1^{tip}/a_1 \approx 40\% \gg da_1^{curve}/a_1$. Therefore, a 40% error is assumed when calculating E' and E'' via adhesive contact models with spherical indenters.

Figure S17(b) shows $a_1$ calculated for an AC160 with the hyperboloid model with adhesion (see section SI2) for radii close to the nominal tip radius (R=7nm) and nominal semi-vertical angle (α=17.5°, taken as the half of the tip nominal back angle). The larger contribution to the error is given by the change in α and leads to a relative error of $da_1^{tip}/a_1 \approx 30\% \gg da_1^{curve}/a_1$. Therefore, a 30% error is assumed when calculating E' and E'' via adhesive contact models with hyperboloid and conical indenters.

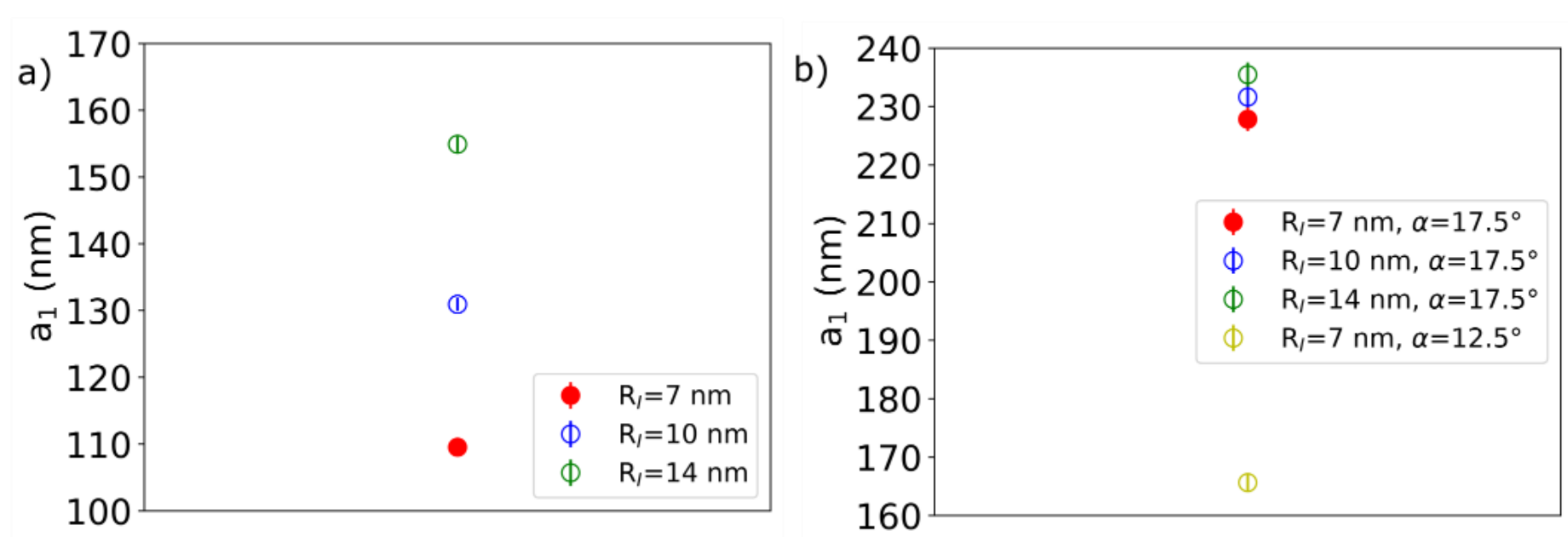

Figure S17. AC160 cantilever contact radius calculated in point 1 ($a_1$) for different contact models and tip sizes: (a) the JKR model with different tip radii, and (b) the adhesive hyperboloid model with different tip radii and semi-vertical angles.